\documentclass[aps,prb,twocolumn]{revtex4-1}
\usepackage{graphicx}
\usepackage{dcolumn}
\usepackage{bm}
\usepackage{color}
\usepackage{amsmath}
\usepackage[export]{adjustbox}

\newcommand{\ket}[1]{\left|#1\right\rangle}

\renewcommand{\Im}{\rm Im}

\newcommand{\up}{\uparrow}
\newcommand{\dn}{\downarrow}

\begin{document}

\title{Renormalization of single-ion magnetic anisotropy in the absence of the Kondo effect}
  
\author{David Jacob} 
\email{david.jacob@ehu.eus}
\affiliation{Nano-Bio Spectroscopy Group and European Theoretical Spectroscopy Facility (ETSF), Departamento de F\'isica de Materiales, Universidad del Pa\'is Vasco UPV/EHU, Av. Tolosa 72, E-20018 San Sebasti\'an, Spain}
\affiliation{IKERBASQUE, Basque Foundation for Science, Mar\'ia D\'iaz de Haro 3, E-48013 Bilbao, Spain}

\date{\today} 

\begin{abstract}
  Inelastic spin flip excitations associated with single-ion magnetic anisotropy
  of quantum spins, can be strongly renormalized by Kondo exchange coupling to the
  conduction electrons in the substrate, as shown recently for the case of Co adatoms on
  CuN$_2$ islands. In this case differential conductance spectra show zero-bias anomalies
  due to a Kondo effect of the doubly degenerate ground state, and finite-bias step features
  due to spin flip excitations.
  Here I consider spin-1 quantum magnets with positive uniaxial anisotropy, where the ground
  state is non-degenerate and hence the Kondo effect does not take place. Nevertheless the
  renormalization of inelastic spin excitations due to exchange coupling by hybridization
  of the quantum spin with the conduction electrons
  still takes place despite the complete absence of the Kondo effect in the ground state. 
  Additionally, I show that away from particle-hole symmetry, charge fluctuations have a similar
  effect to Kondo exchange coupling, leading to the renormalization of spin flip excitations.
  However, in contrast to the renormalization by Kondo exchange, charge fluctuations lead to
  asymmetric spectra, which for strong charge fluctuations can mimic Fano behavior.
\end{abstract}

\maketitle

\section{Introduction}

The interaction of a nanoscale quantum magnet, realized e.g. by single magnetic atoms or molecules
deposited on substrates or coupled to conducting electrodes, with the environment can have a
strong effect on its properties.
On the one hand, the interaction with the conduction electrons of a nearby electrode can
give rise to the Kondo effect, which leads to the screening of the magnetic moment of the quantum
magnet (see e.g. the book by Hewson\cite{Hewson:book:1997} and references therein).
On the other hand the crystal field of the environment of a quantum magnet in conjunction
with spin-orbit coupling (SOC) gives rise to magnetic anisotropy (MA), which generally leads to a
stabilization of the magnetic moment of the nanoscale magnet.\cite{Gatteschi:book:2006}

These two antagonistic effects can be observed by differential conductance spectroscopy in Scanning Tunneling Microscope
(STM) experiments of magnetic adatoms and molecules deposited on conducting and insulating
substrates,\cite{Madhavan:Science:1998,Li:PRL:1998,Zhao:Science:2005,Iancu:NL:2006,Hirjibehedin:Science:2007}
and in molecular junctions:\cite{Yu:PRL:2005,Parks:PRL:2007,Calvo:Nature:2009,Parks:Science:2010}
The Kondo effect, which leads to the formation of a sharp resonance in the electronic spectrum at the Fermi
level, called the Abrikosov-Suhl or Kondo resonance, is signaled by a corresponding zero-bias anomaly in the
$dI/dV$ spectra.\cite{Madhavan:Science:1998,Schiller:PRB:2000,Ujsaghy:PRL:2000}
Magnetic anisotropy on the other hand leads to spin flip excitations from the ground states,
which are signaled by steps in the $dI/dV$-spectra at finite bias voltages corresponding to the excitation energies
of the quantum magnet.\cite{Hirjibehedin:Science:2007,Fernandez-Rossier:PRL:2009}
Both effects, i.e. Kondo peak plus spin flip excitation steps in the 
$dI/dV$, have also been observed simultaneously in the same system, namely in Co adatoms on CuN islands.\cite{Otte:NatPhys:2008}
On the other hand, sometimes the same molecule [iron(II) phthalocyanine (FePc)] can show the Kondo effect on
one substrate [Au(111)]\cite{Tsukahara:PRL:2011} and spin flip excitations on another [oxidized Cu(110)]~\cite{Tsukahara:PRL:2009}.
More recently, it has been shown experimentally and theoretically that the Kondo exchange coupling between a quantum
spin and the conduction electrons in the substrate, also leads to the renormalization of spin excitation energies
associated with the magnetic anisotropy of the quantum magnet.\cite{Oberg:NatNano:2014}
Relatedly, very recently it has been found that by lifting a spin-1 molecule (FePc) from a
Au surface, and thereby reducing the Kondo exchange coupling with the conduction electrons in the surface, the molecule
makes a transition from a Kondo screened state, showing a zero-bias anomaly to an essentially unscreened state, showing inelastic spin
flip excitations at finite bias.\cite{Hiraoka:NatComm:2017}

Here I show, that even in the complete absence of the Kondo effect, i.e. for an integer quantum spin where the
degeneracy of the ground state is completely lifted by a positive uniaxial magnetic anisotropy, the exchange
coupling to the conduction electrons still leads to the renormalization of the MA related spin flip excitations
of the quantum magnet. To this end a multi-orbital Anderson model for the adatom coupled to the substrate
and subject to magnetic anisotropy is solved within the one-crossing approximation (OCA).
While the inelastic steps as well as the Kondo features can also be described using
a Kondo Hamiltonian where the atomic spin is described with a single-ion quantized spin interacting, 
via exchange, with the conduction electrons of the surface,\cite{Zitko:PRB:2008,Fernandez-Rossier:PRL:2009,Lorente:PRL:2009,
  Zitko:NJP:2010,Hurley:PRB:2011,Oberg:NatNano:2014,Delgado:SS:2014,Ternes:NJP:2015}
this approach permits to include atomic charge fluctuations that are effectively frozen in the Kondo model. 
This can be important since often charge is not quantized in magnetic adatom systems as density functional theory (DFT)
calculations show.\cite{Ferron:PRB:2015,Panda:PRB:2016}
Indeed below I show that charge fluctuations have a similar effect on the electronic spectra and related $dI/dV$,
leading to the renormalization of the spin flip excitations. But in contrast to the renormalization by
exchange coupling, the spectra become increasingly asymmetric as the charge fluctuations grow. Interestingly,
for very strong charge fluctuations this leads to features in the spectra that mimic Fano lineshapes usually
associated with zero-bias anomalies, such as the Kondo resonance.

This paper is organized as follows: In Sec.~\ref{sec:model} the model describing a quantum spin
subject to magnetic anisotropy coupled to a conducting electrode is introduced.
In Sec.~\ref{sec:results} results for an $S=1$ quantum spin with positive uniaxial MA are presented.
Additionally also the effect of charge fluctuations for a $S=3/2$ quantum spin is shown.
Sec.~\ref{sec:conclusions} I conclude with a general discussion of the results.

\begin{figure*}
  \begin{center}
    \includegraphics[width=\linewidth]{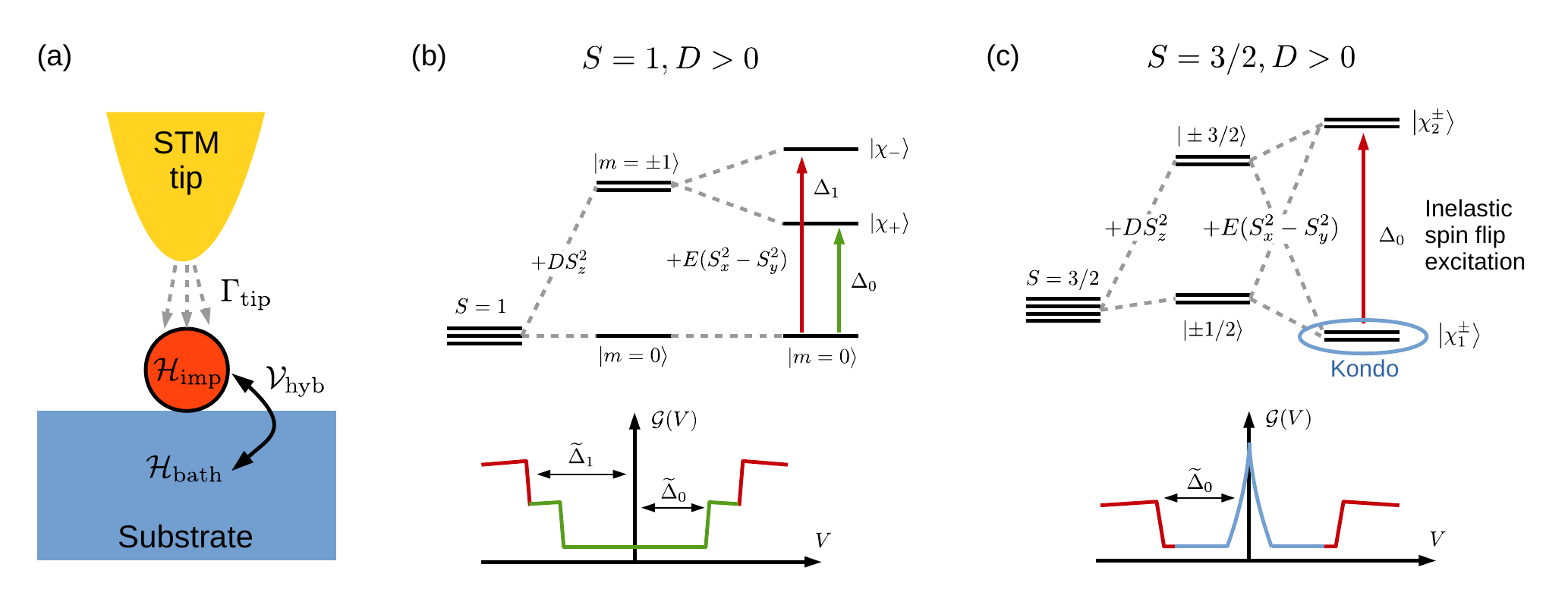}
  \end{center}
  \caption{
    \label{fig:model}
    (a) Schematic model of experimental setup for measuring
    excitation spectra of a magnetic adatom on a surface
    with an STM tip.
    (b) Splitting of the spin $S=1$ GS triplet by the MA term in eq. (\ref{eq:imp})
    for positive uniaxial anisotropy $D>0$ and finite in-plane anisotropy $0<E<D/3$, and
    resulting differential conductance $\mathcal{G}(V)$
    spectrum (bottom), showing two inelastic steps features for each bias direction 
    corresponding to the two (renormalized) spin excitations $\widetilde\Delta_0$ (green) 
    and $\widetilde\Delta_1$ (red). In contrast to the $S=3/2$ case (c) the
    Kondo effect cannot take place as the GS is non-degenerate.
    (c) Splitting of the spin $S=3/2$ GS quadruplet by the MA term in eq. (\ref{eq:imp})
    into two doublets. The GS doublet $\left\{\ket{\chi^\pm_1}\right\}$ can give rise to the 
    Kondo feature (blue) in the differential conductance $\mathcal{G}(V)$ spectrum (bottom)
    while the spin excitation to the $\left\{\ket{\chi^\pm_2}\right\}$ doublet gives rise
    to inelastic step features (green) at the (renormalized) excitation energy $\widetilde\Delta_0$.
  }
\end{figure*}

\section{Model and Method}
\label{sec:model}

Let us consider tunneling between an STM tip and a magnetic adatom coupled to a surface
as shown in Fig.~\ref{fig:model}(a).
Assuming weak coupling to the STM tip,
the low-bias conductance can be directly related to the 
adatom many-body spectral function $\rho_\alpha(\omega)$:\cite{Korytar:PRB:2012} 
\begin{equation}
  \mathcal{G}(V) = \frac{2e^2}{\hbar} \sum_{\alpha} \Gamma^{\rm tip}_{\alpha} \rho_{\alpha}(eV) 
  \label{GdeV}
\end{equation}
where $\Gamma^{\rm tip}_{\alpha}=\pi|{V_{\alpha}^{\rm tip}}|^2\rho_{\rm tip}$ 
is the (weak) tunneling rate of electrons between the adatom orbitals $\alpha$ and  the STM tip,
and it is assumed the DOS of the STM tip $\rho_{\rm tip}$ is energy independent around the 
Fermi level. 
Note however, that in general different orbitals couple differently 
to the STM tip so that the contribution of the individual channels 
to the total conductance may differ.
Direct tunneling into surface states is neglected in eq.~(\ref{GdeV}). This is 
a good approximation when the magnetic atoms are separated from the metallic surface
by a decoupling insulating layer, such as Cu$_2$N/Cu(100),\cite{Hirjibehedin:Science:2007,Oberg:NatNano:2014,Otte:NatPhys:2008}
CuO/Cu\cite{Tsukahara:PRL:2009} and h-BN/Rh(111)\cite{Jacobson:NatComm:2015}. This approximation does not
capture the Fano interference effect relevant\cite{Madhavan:Science:1998} when the tip-atom 
channel interferes with the direct tip-surface tunneling path.\cite{Ujsaghy:PRL:2000}
In this case a more complex modeling of the tunneling process would be necessary.\cite{Jacob:JPCM:2015,Baruselli:PRB:2015,Frank:PRB:2015,Dang:PRB:2016,Choi:JCP:2017,Droghetti:PRB:2017}

The magnetic atom on the surface is described by a multi-orbital Anderson model,
\begin{equation}
  {\cal H} = \mathcal{H}_{\rm imp} + {\mathcal H}_{\rm bath} + {\cal V}_{\rm hyb}
  \label{eq:AIM}
\end{equation}
where the Hamiltonian of the Anderson impurity site $\mathcal{H}_{\rm imp}$ 
describes the strongly interacting $3d$-levels that yield the spin of 
the magnetic atom, and includes a term that accounts for magnetic anisotropy:
\begin{eqnarray}
  {\cal H}_{\rm imp} &=& \epsilon_d \hat{N}_d 
  + U \sum_{\alpha} \hat{n}_{\alpha\up} \, \hat{n}_{\alpha\dn} 
  + U^\prime \sum_{{\alpha,\alpha^\prime}\atop{\alpha\neq\alpha^\prime}} \hat{n}_{\alpha} \, \hat{n}_{\alpha^\prime} 
  \nonumber\\
  &-& J_{\rm H} \sum_{{\alpha,\alpha^\prime}\atop{\alpha\neq\alpha^\prime}} \vec{S}_\alpha \cdot \vec{S}_{\alpha^\prime} 
  + D \hat{S}_z^2 + E (\hat{S}_x^2 - \hat{S}_y^2 )
  \label{eq:imp}
\end{eqnarray}
$\epsilon_d$ are the single-particle energies of the $d$-levels,
$\hat{N}_d=\sum_{\alpha,\sigma}\hat{n}_{\alpha\sigma}$ is the number operator
for all $d$-levels $\alpha=1,\ldots,M$, $\hat{n}_{\alpha\sigma}=d_{\alpha\sigma}^\dagger d_{\alpha\sigma}$ 
is the number operator of an individual $d$-level $\alpha$ with spin $\sigma$, 
$U$ is the intra-orbital, $U^\prime$ the inter-orbital Coulomb repulsion, $J_{\rm H}$ the Hund's coupling, 
and $\vec{S}_\alpha$ measures the total spin of an individual $d$-level $\alpha$, 
i.e. $\vec{S}_\alpha= \sum_{\sigma\sigma^\prime} d_{\alpha\sigma}^\dagger \vec\tau_{\sigma\sigma^\prime} d_{\alpha\sigma^\prime}$.

The crystal field splitting of the $d$-levels 
together with the spin-orbit coupling (SOC) gives rise \cite{Abragam:book:2012} 
to magnetic anisotropy (MA) which is taken into account
by the effective spin Hamiltonian given by the last term of (\ref{eq:imp}) where 
$D$ is the uniaxial anisotropy and $E$ the in-plane anisotropy.\cite{Gatteschi:book:2006}
For a ground state (GS) with integer spin $S$, and for finite uniaxial anisotropy 
$D\ne0$ and in-plane anisotropy $E\ne0$, the $(2S+1)$ degeneracy of the GS multiplet 
is completely lifted\cite{Abragam:book:2012,Hirjibehedin:Science:2007}, as illustrated for $S=1$ in Fig.~\ref{fig:model}(b).
For $D>0$ the $\ket{m_z=0}$ state becomes the GS and the $\ket{m_z=\pm1}$ states
an excited doublet. A finite in-plane anisotropy $E>0$ allows for \emph{quantum tunneling}
between both spin directions thus lifting the degeneracy of the excited doublet, 
which becomes split by $2E$, leading to the bare excitation energies $\Delta_0=D-E$
and $\Delta_1=D+E$. The quantum states of the split doublet are thus linear combinations 
$\ket{\chi_\pm}\sim\ket{m_z=+1}\pm\ket{m_z=-1}$.

In contrast, for half-integer spin the degeneracy of the GS multiplet is never
completely lifted by the magnetic anisotropy, as %%shown in App.~\ref{App:SpinHamiltonian} and
illustrated in Fig.~\ref{fig:model}(c) for $S=3/2$.
For $D>0$ and $E=0$ the GS quadruplet splits into two doublets, with the states $\ket{m_z=\pm1/2}$
forming the GS and the $\ket{m_z=\pm3/2}$ states forming the excited doublet. 
A finite in-plane anisotropy $E$ leads to the mixing of states of the GS doublet 
and excited doublet, but the double degeneracy of the GS and excited state is conserved.

The second term in (\ref{eq:AIM}) describes the conduction electron bath in the surface:
\begin{equation}
  {\cal H}_{\rm bath}=\sum_{k,\alpha,\sigma} \varepsilon_{k\alpha} c_{k\alpha\sigma}^{\dagger}c_{k\alpha\sigma}
\end{equation}
The third term in (\ref{eq:AIM}) is the so-called hybridization term which describes the 
coupling between the impurity and the conduction electron bath:
\begin{equation}
  \label{eq:Vhyb}
  {\cal V}_{\rm hyb} = \sum_{k,\alpha,\sigma} V_{k\alpha} (c_{k\alpha\sigma}^\dagger d_{\alpha\sigma} + d_{\alpha\sigma}^\dagger c_{k\alpha\sigma})
\end{equation}
Integrating out the bath degrees of freedom one obtains the so-called hybridization function:
\begin{equation}
  \label{eq:hybfunc}
  \Delta_\alpha^{\rm hyb}(\omega)=\sum_{k}\frac{|V_{k\alpha}|^2}{\omega+\mu-\varepsilon_{k\alpha}+i\eta}
\end{equation}
Its (negative) imaginary part $\Gamma_\alpha(\omega)=-\Im\,\Delta_\alpha^{\rm hyb}(\omega)$ describes the single-particle broadening 
of individual impurity levels $\alpha$ due to the coupling to the conduction electrons.

The Anderson model (\ref{eq:AIM}) is then solved within the one-crossing approximation (OCA)\cite{Pruschke:ZPB:1989,Haule:PRB:2001,Haule:PRB:2010},
as described in more detail previous work.\cite{Jacob:EPJB:2016}
OCA consists in a diagrammatic expansion of the propagators $G_n(\omega)$ associated with the
many-body eigenstates $\ket{n}$ of the \emph{isolated} impurity Hamiltonian (\ref{eq:imp}) in terms of
the hybridization function $\Delta_\alpha^{\rm hyb}(\omega)$, summing only a subset of diagrams
(only those where conduction electron lines cross at most once) to infinite order. 
The spectral function $\rho_\alpha(\omega)$ entering equation 
(\ref{GdeV}) for the $dI/dV$ is then obtained from convolutions of the propagators $G_n(\omega)$.

\section{Results}
\label{sec:results}

In order to model a spin-1 adatom or molecule coupled to conducting electrodes, the two-orbital
Anderson model at and around half-filling ($N_d=2$) is now studied. The Hund's rule coupling $J_{\rm H}$ then leads
to an $S=1$ ground state for the isolated impurity $\mathcal{H}_{\rm imp}$ that is split by the 
magnetic anisotropy term in (\ref{eq:imp}) with positive uniaxial anisotropy $D>0$ into the GS singlet
$\ket{m_z=0}$ and the two excited states $\ket{\chi_\pm}$. %%as described in App.~\ref{App:SpinHamiltonian}.
Furthermore the hybridization function is assumed to be constant and equal for both orbitals, $\Delta_\alpha=-i\Gamma$.
To achieve exactly the half-filled (i.e. the ph symmetric) case, the impurity levels have to be tuned to the energy
\begin{equation}
  \label{eq:ed_ph}
  \epsilon_d^\ast = -\frac{U}{2} -\left(U^\prime-\frac{J_{\rm H}}{2}\right)(N_d-1)
\end{equation}
where $N_d$ is the number of electrons.
For all cases the same interaction parameters, $U=3.5$eV, $U^\prime=2.5$eV, and $J_{\rm H}=0.5$eV,
are considered which leads to $\epsilon_d^\ast=4$eV at half-filling ($N_d=2$).

\begin{figure}
  \begin{center}
    \includegraphics[width=0.9\linewidth]{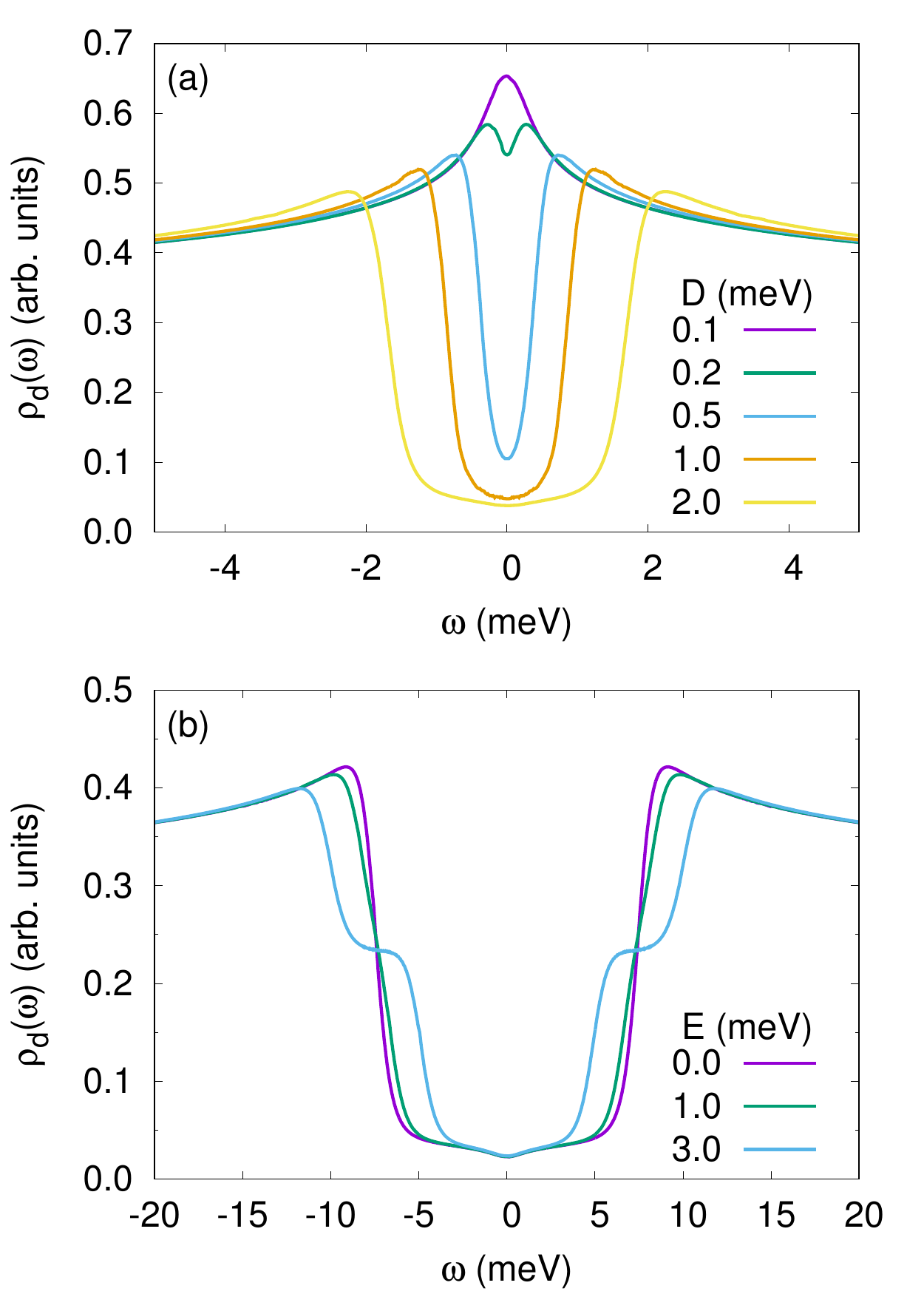}
  \end{center}
  \caption{
    \label{fig:spin1_MA}
    Effect of magnetic anisotropy on an $S=1$ quantum spin. 
    Intra- and inter-orbital Coulomb repulsion are $U=3.5$eV and $U^\prime=2.5$eV, respectively, Hund's rule
    coupling is $J_{\rm H}=0.5$eV and the $d$-level energy is $\epsilon_d=-4$eV (ph symmetric case).
    (a) Effect of increasing uniaxial anisotropy $D$ on the electronic spectrum for $\Gamma/\pi=50$meV and temperature $kT=0.1$meV.
    (b) Effect of increasing in-plane anisotropy $E$ on the electronic spectrum for $D=8.7$meV and $\Gamma/\pi=50$meV and temperature $kT=0.4$meV.
  }
\end{figure}

In Fig.~\ref{fig:spin1_MA} spectra for a spin $S=1$ adatom or molecule with positive uniaxial anisotropy
$D>0$ are shown. Fig.~\ref{fig:spin1_MA}(a) shows how the uniaxial anisotropy $D$ splits the Kondo
peak of the spin-1 Kondo effect at $D=0$. Analogously to the splitting of the Kondo peak for the
spin-1/2 Kondo effect in a magnetic field, the split Kondo peak for small values of $D$ gradually
develops into two step features (one for each direction of energy) for larger values of $D$.
These steps correspond to spin excitations from the $m_z=0$ GS to the excited doublet $m_z=\pm1$.
Such spectra have recently been measured in STM experiments of spin-1 molecules, namely
Fe porphyrins on Au\cite{Karan:NL:2018,Rubio-Verdu::2017} or nickelocene on Cu\cite{Ormaza:NatComm:2017}.
Note that in contrast to the case of a spin-1 quantum magnets with negative uniaxial anisotropy, $D<0$,\cite{Jacob:EPJB:2016}
a Kondo peak at zero energy is absent now from the spectrum, due to the lack of GS degeneracy.
Also note that the simpler noncrossing approximation (NCA) leads to a spurious Kondo-like artifact
in the spectrum in this case. Apparently, the vertex corrections present in the OCA approach remedy this problem
of the NCA approach (see Appendix \ref{app:nca_artifact}).
As can be seen in Fig.~\ref{fig:spin1_MA}(b), the introduction of in-plane anisotropy $E$, leads to the appearance
of two new step features in the spectrum, one for each direction of energy, due to the splitting of the excited
doublet $m_z=\pm1$, which lead to different excitation energies $\Delta_0$ and $\Delta_1$ from the GS to the two
now split excited spin states, as shown schematically in Fig.~\ref{fig:model}.
This double step structure has been observed in the $dI/dV$ spectra of numerous spin-1 quantum magnets such as
FePc on oxidized Cu(110),\cite{Tsukahara:PRL:2009} or hydrogenated Fe atoms on hexagonal boron nitride.\cite{Jacobson:NatComm:2015}

\begin{figure}
  \begin{center}
    \includegraphics[width=0.9\linewidth]{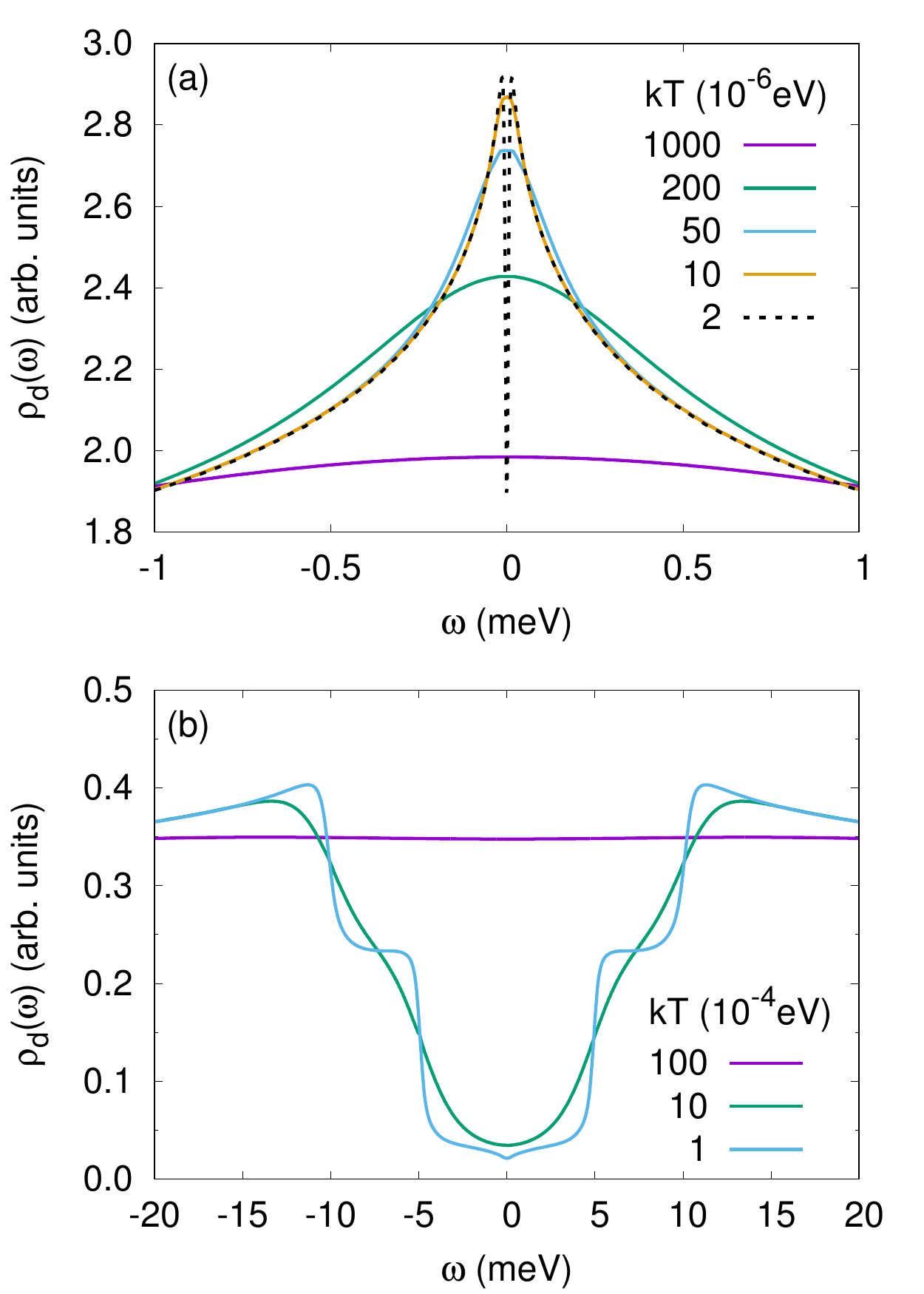}
  \end{center}
  \caption{
    \label{fig:spin1_temp}
    Temperature dependence of spectra for an $S=1$ quantum spin with (a) very weak uniaxial anisotropy $D=0.1$meV and $\Gamma/\pi=150$meV,
    and (b) substantial uniaxial and in-plane anisotropy ($D=8.7$meV and $E=3$meV) and $\Gamma/\pi=50$meV.
    Same impurity parameters ($U$, $U^\prime$,$J_{\rm H}$,$\epsilon_d$) as in Fig.~\ref{fig:spin1_MA}.
  }
\end{figure}

Next the temperature dependence of the spectra in Fig.~\ref{fig:spin1_temp} is investigated,
specifically the temperature evolution for two cases: (a) for very weak uniaxial
anisotropy ($D=0.1$meV and $E=0$) and relatively strong hybridization ($\Gamma/\pi=150$meV),
and (b) strong uniaxial and in-plane anisotropy and relatively weak hybridization ($\Gamma/\pi=50$meV).
First, in the case of very weak uniaxial MA [Fig.~\ref{fig:spin1_temp}(a)] upon lowering the
temperature a Kondo resonance forms at the Fermi level, associated with a spin-1 Kondo effect,
fully screened by the conduction electrons (there are two screening channels).
Interestingly, even for this very weak anisotropy of $D=0.1$meV the Kondo peak eventually
splits at low enough temperatures, even though the Kondo scale for the spin-1 Kondo effect,
$kT_{\rm K}\sim2$meV (estimated from the half-width of the Kondo peak in the absence of anisotropy)
is now considerably larger than the $D$. Na\"ively, one would expect the Kondo screening
to completely overcome the marginal lifting of the spin triplet degeneracy by the weak MA.
However, scaling arguments as well as NRG calculations of the anisotropic one-channel and
multi-channel Kondo models show\cite{Schiller:PRB:2008,Zitko:PRB:2008} that a magnetic anisotropy
term $\sim{DS_z^2}$ is always reinforced at lower energies (i.e. low frequencies $\omega$ and low
temperatures $T$). Hence for integer spins and $D>0$ in the limit $T\rightarrow0$ the spectral
density always acquires a dip around the Fermi level.
In Fig.~\ref{fig:spin1_temp}(b) the temperature dependence for a spin-1 system with both appreciable
uniaxial and inplane MA is shown. Now on lowering the temperature a dip forms at first instead of the
Kondo peak. On further decrease of the temperature the dip develops into the well known step features
associated with inelastic spin flip excitations. Simultaneously, an ``up bending'' of the spectral
density around the step features occurs when the temperature is lowered, giving rise to the familiar
triangular shapes of the step features well known from STM spectroscopy of magnetic adatoms and
molecules.

\begin{figure}
  \begin{center}
    \includegraphics[width=0.9\linewidth]{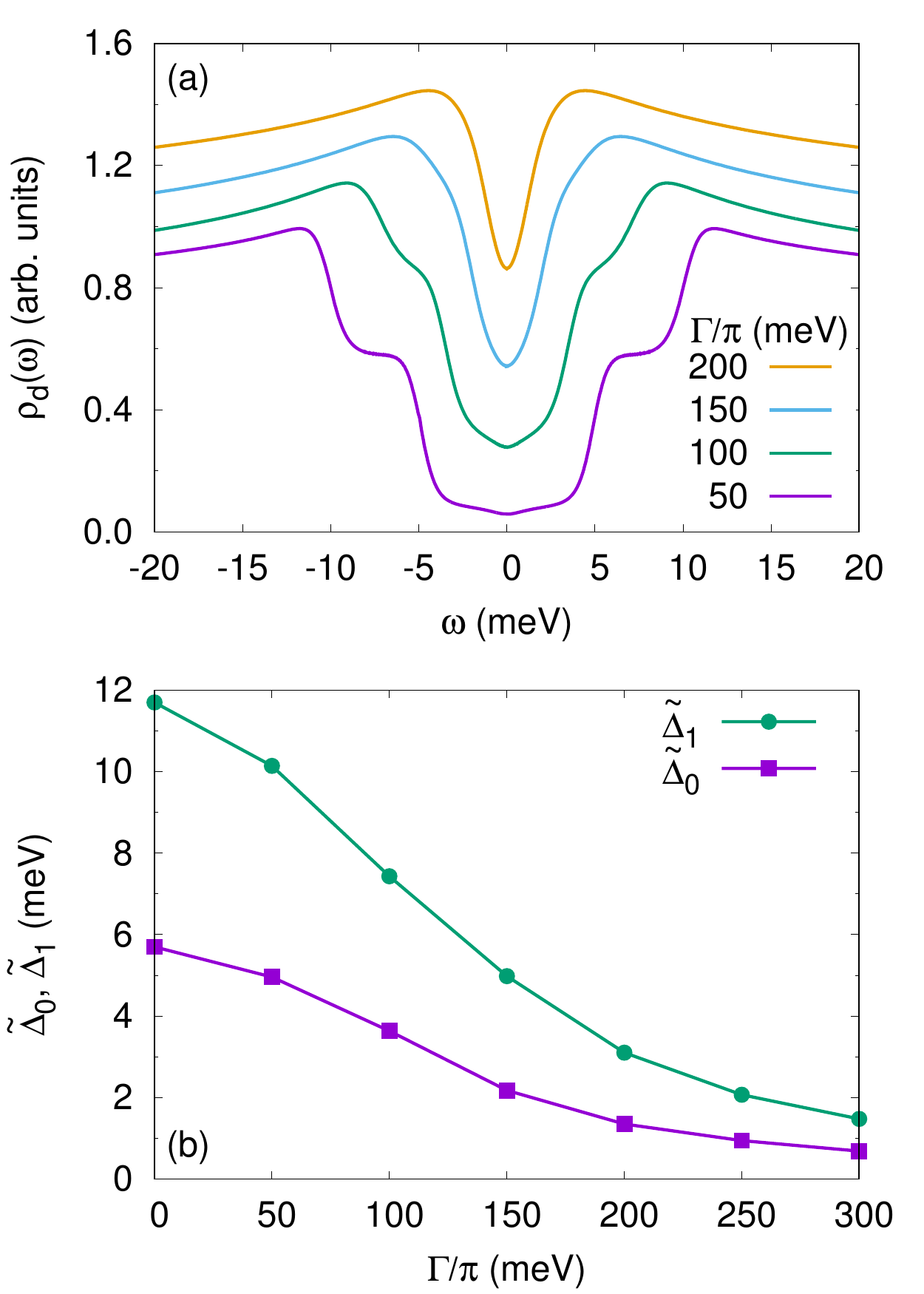}
  \end{center}
  \caption{
    \label{fig:spin1_hyb}
    Effect of hybridization $\Gamma$ on an $S=1$ quantum spin with positive uniaxial anisotropy $D$.
    Same impurity parameters ($U$, $U^\prime$,$J_{\rm H}$,$\epsilon_d$) as in Fig.~\ref{fig:spin1_MA}.
    (a) Effect of increasing hybridization $\Gamma$ on the electronic spectrum for $D=8.7$meV and $E=3$meV, $kT=0.4$meV.
    The spectra have been normalized to their maximum value in the considered energy window and shifted for
    better visibility.
    (b) Effective spin excitation energies $\widetilde\Delta_0$ and $\widetilde\Delta_1$ as a function of 
    $\Gamma$ for $D=8.7$meV and $E=3$meV.
  }
\end{figure}

Let us now study the effect of the exchange coupling with the conduction electrons on the spin excitations.
According to Schrieffer-Wolff \cite{Schrieffer:PR:1966} the Kondo exchange coupling $J_K$ of an impurity spin is directly proportional
to the hybridization $\Gamma$ of the impurity with the conduction electrons, $J_K\sim\Gamma/\Delta{U}$ where
$\Delta{U}$ is the charging energy of the impurity shell.
Fig.~\ref{fig:spin1_hyb}(a) shows how the low energy spectra change as the hybridization with the
conduction electrons $\Gamma$ increases, where the uniaxial anisotropy was chosen to model that of FePc
molecules in the gas phase, $D\sim8.7$meV, but with finite inplane anisotropy, $E=3$meV
to account for symmetry breaking on the substrate.\cite{Tsukahara:PRL:2009}
One can see that the effect of increasing the coupling to the conduction electrons is quite strong.
As $\Gamma$ increases, the step features become broader due to the broadening of the electronic levels
by the coupling to the conduction electrons, and importantly move towards the center, i.e. to lower energies,
signaling the renormalization of the corresponding spin flip excitations by the exchange coupling with the
conduction electrons.
Fig.~\ref{fig:spin1_hyb}(b) shows how the \emph{effective} excitation energies\footnote{The excitations energies $\tilde\Delta_0$ and
  $\tilde\Delta_1$ are most conveniently extracted from the pseudo particle spectra associated to the corresponding excited many body
  states which show sharp peak features at the positions of the spin flip steps in the real electron spectrum, as shown in previous work.\cite{Jacob:EPJB:2016}}
$\tilde\Delta_0$ and $\tilde\Delta_1$ shrink as the hybridization with the conduction electrons $\Gamma$, and correspondingly the
Kondo exchange, increases. Note that even for very large $\Gamma$, the Kondo exchange does not overcome the
magnetic anisotropy, due to the above discussed reinforcement of the magnetic anisotropy at lower energies according to
the scaling arguments and NRG results of \v{Z}itko {\it et al.}~\cite{Zitko:PRB:2008}
Also note that for strong hybridization, $\Gamma/\pi\ge200$meV, the inner and outer inelastic steps become close in energy
and thus are not resolved at the temperature considered in Fig.~\ref{fig:spin1_hyb}(a). Lowering the temperature, the individual
steps can be resolved (not shown), similar to the case shown in Fig.~\ref{fig:spin1_temp}(b), although the steps are much less pronounced now.

\begin{figure}
  \begin{center}
    \includegraphics[width=0.9\linewidth]{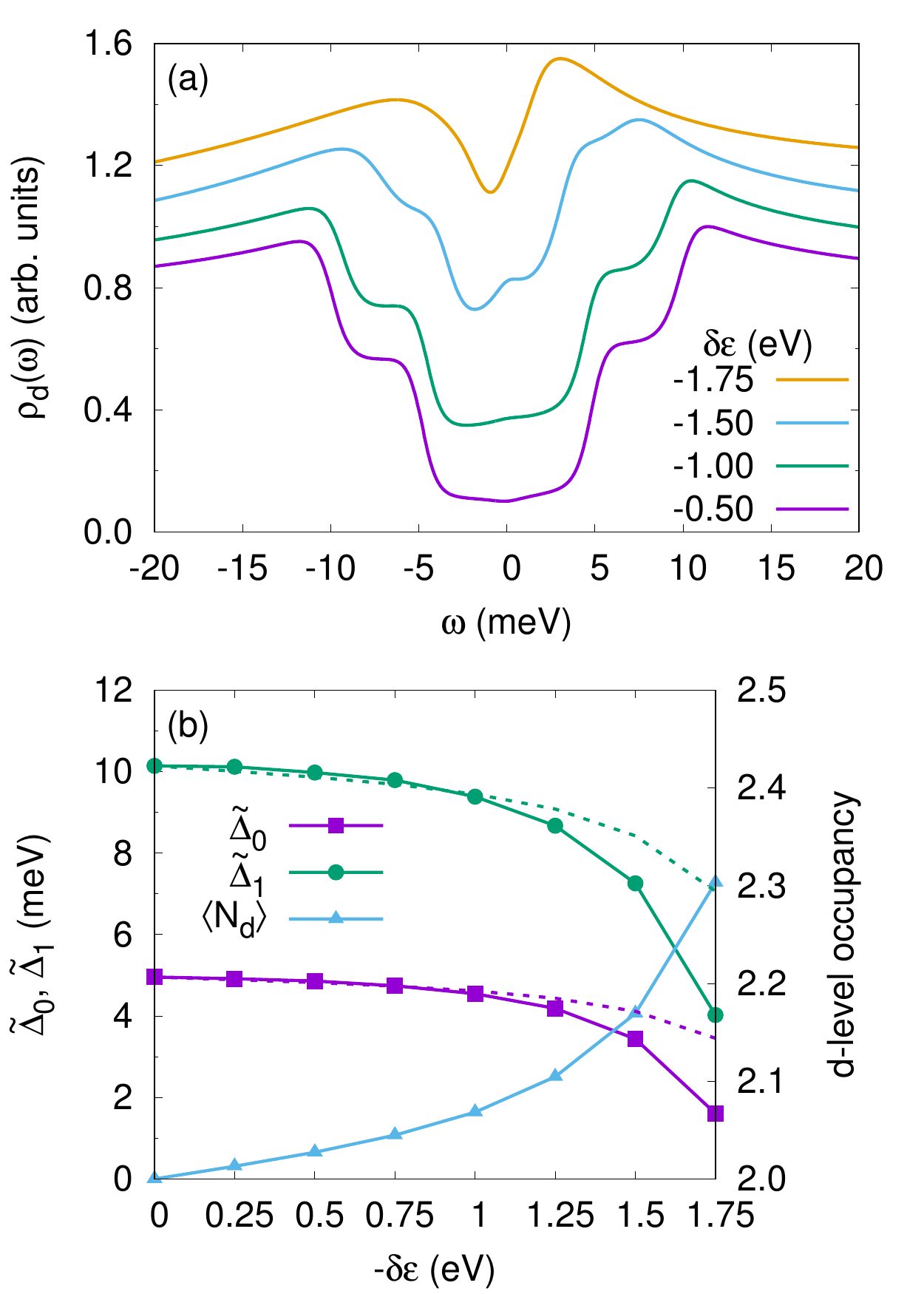}
  \end{center}
  \caption{
    \label{fig:spin1_Ed}
    Effect of valence fluctuations on an $S=1$ quantum spin with positive uniaxial anisotropy $D$.
    Same impurity parameters ($U$, $U^\prime$,$J_{\rm H}$) as in Fig.~\ref{fig:spin1_MA}, but with
    $d$-level energy $\epsilon_d$ detuned from ph symmetry.
    (a) Effect of charge fluctuations induced by detuning $\delta\epsilon$ of $d$-level energy
    away from ph symmetry on the electronic spectrum for $D=8.7$meV, $E=3$meV and $\Gamma/\pi=50$meV, $kT=0.4$meV.
    The spectra have been normalized to their maximum value in the considered energy window and shifted for
    better visibility. 
    (b) Effective spin excitation energies $\widetilde\Delta_0$ and $\widetilde\Delta_1$ as a function of
    the $d$-level detuning from ph symmetry $\delta\epsilon$. Dashed lines show the renormalization
    only due to the reduction of the spin by the charge fluctuations according to (\ref{eq:Delta}).
  }
\end{figure}

Let us now study the effect of valence fluctuations on the spectra and spin excitations of a spin-1 quantum magnet.
Detuning of the impurity levels away from exact ph symmetry (i.e. half-filling), $\epsilon_d=\epsilon_d^\ast+\delta\epsilon$,
leads to the impurity valence fluctuating between $N=2$ and $N+1=3$ electrons (for $\delta\epsilon<0$).
Accordingly, the charge of the impurity shell takes a fractional value between $N$ and $N+1$:
\begin{equation}
  \label{eq:charge}
  N_d=(1-\alpha)N+\alpha(N+1)=N+\alpha.
\end{equation}
Fig.~\ref{fig:spin1_Ed}(a) shows the effect of detuning from exact ph symmetry on the spectrum of an $S=1$ quantum spin
with both uniaxial and in-plane MA (same as in Fig.~\ref{fig:spin1_hyb}). Similar to the effect of increasing hybridization
(Fig.~\ref{fig:spin1_hyb}), the step features move to lower energies, as the detuning $|\delta\epsilon|$ and thus the
concomitant valence fluctuations grow [see blue curve showing impurity charge $N_d$ in Fig.~\ref{fig:spin1_Ed}(b)],
indicating a renormalization of the associated spin excitation energies by the valence fluctuations.
But in contrast to the renormalization by Kondo exchange the spectra become increasingly
asymmetric as $|\delta\epsilon|$ increases, due to the lifting of the ph symmetry.
Note that asymmetric $dI/dV$ spectra are often observed experimentally.~\cite{Karan:NL:2018,Rubio-Verdu::2017}
The increasing asymmetry culminates in the formation of a Fano-like lineshape for strong detuning from ph symmetry
($\delta\epsilon\sim1.8$eV). Fano lineshapes are often observed in STM spectroscopy of magnetic
impurities on conducting substrates, where it is usually accepted as a fingerprint for
the Kondo effect resulting from quantum interference of different tunneling paths from the tip
to the substrate, one(s) going through impurity orbital(s) bearing a Kondo peak,
and others through orbitals without, i.e. simply broadened by the substrate, or
going directly into the substrate.\cite{Frank:PRB:2015}
Here in contrast, the Fano-like feature in the spectral function emerges due to the combined effect of
strong ph asymmetry of the spectrum and the two spin flip steps moving to lower energies.
Hence a Fano lineshape measured by STM spectroscopy of magnetic atoms or molecules on conducting substrates
does not necessarily indicate the occurrence of the Kondo effect.

Fig.~\ref{fig:spin1_Ed}(b) shows the effective spin excitation energies $\tilde\Delta_0$ and $\tilde\Delta_1$
as well as the impurity charge $\langle\hat{N}_d\rangle$ as a function of the detuning $\delta\epsilon$.
Clearly, the decrease of $\tilde\Delta_0$ and $\tilde\Delta_1$ as $|\delta\epsilon|$ grows, is linked to the
corresponding growth of valence fluctuations, indicated by the increasing deviation of $\langle\hat{N}_d\rangle$
from integer occupation of the impurity shell. Part of the reduction of $\tilde\Delta_0$ and $\tilde\Delta_1$
actually stems from the reduction of the spin of the impurity shell due to the deviation from half-filling,
according to
\begin{equation}
  \langle S^2 \rangle_{N_d} = (1-\alpha) \langle S^2 \rangle_{N=2} + \alpha \langle S^2 \rangle_{N+1=3},
\end{equation}
leading to a corresponding reduction in the spin excitation energies, according to
\footnote{The $N+1=3$ subspace has spin-1/2 (one hole) and thus does not give rise to MA.}
\begin{equation}
  \label{eq:Delta}
  \tilde\Delta_n(\Gamma,N_d) = (1-\alpha) \tilde\Delta_n(\Gamma,N) \mbox{ with } n=0,1
\end{equation}
where $\alpha$ is determined by (\ref{eq:charge}) and the actual charge of the impurity shell $\langle\hat{N}_d\rangle$.
The dashed lines in Fig.~\ref{fig:spin1_Ed} show the reduction of the spin excitation energies
due to the reduction of the spin alone, according to (\ref{eq:Delta}). It can be seen that only for
small deviations from ph symmetry, and thus for weak valence fluctuations, does the reduction of the
effective spin due to charge fluctuations account for the reduction in the spin excitation energies entirely. 
For larger deviations and stronger valence fluctuations, the reduction of the spin only accounts for
part of the reduction in the spin excitation energies. The other part must then originate from the
renormalization due to quantum fluctuations of the charge of the impurity shell, similar to the
renormalization of the spin excitation energies by quantum fluctuations of the spin by the Kondo
exchange coupling.

\begin{figure}
  \begin{center}
    \includegraphics[width=0.9\linewidth]{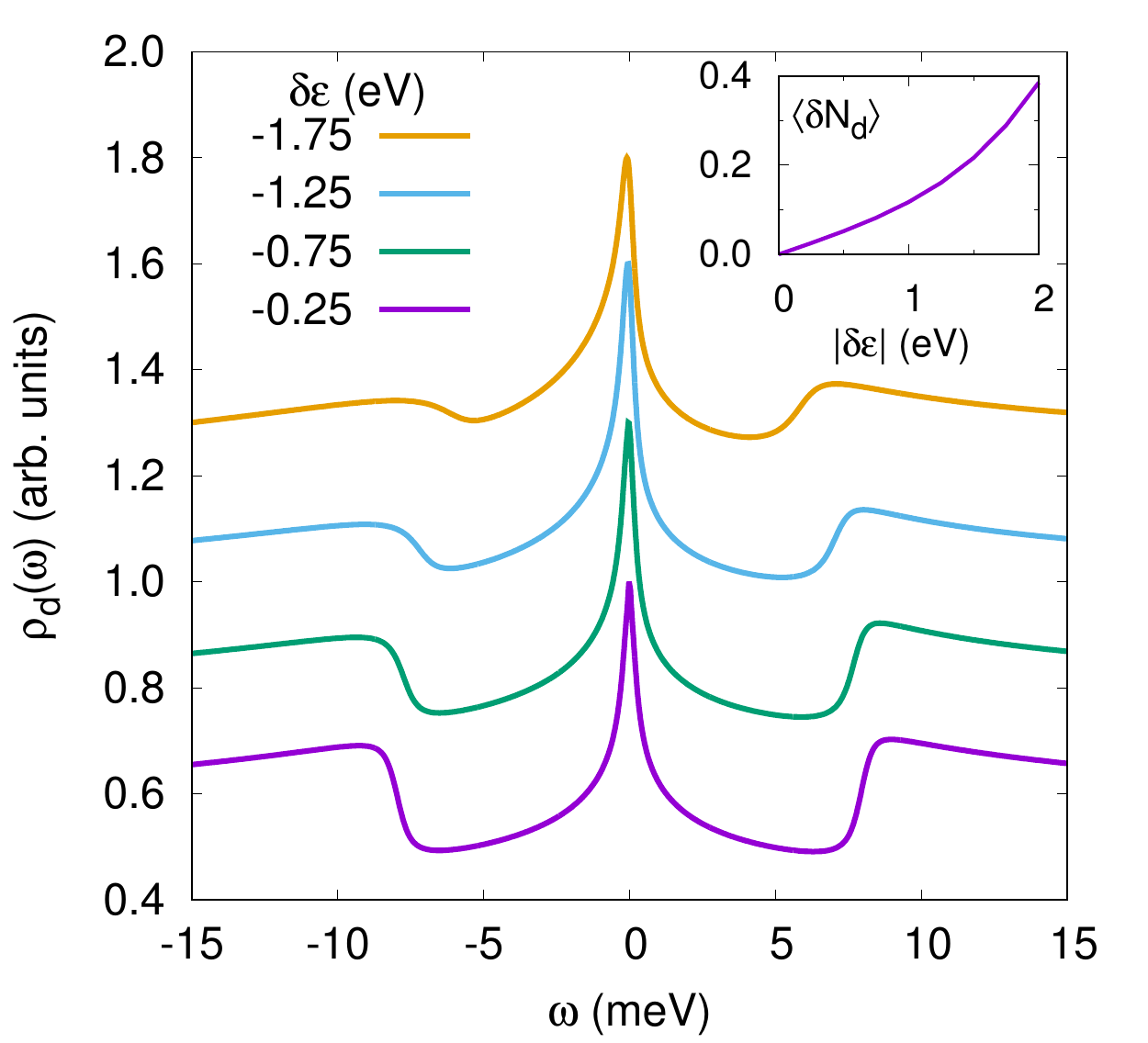}
  \end{center}
  \caption{
    \label{fig:spin3/2}
    Effect of valence fluctuations on an $S=3/2$ quantum spin with uniaxial anisotropy $D=5$meV,
    hybridization $\Gamma/\pi=100$meV, direct Coulomb repulsion $U=3.5$eV, $U^\prime=2.5$eV and
    Hund's rule coupling $J_{\rm H}=0.5$eV. The $d$-level energy is $\epsilon_d=-6.25$eV for the
    ph symmetric case. $\delta\epsilon$ is the detuning of the impurity level $\epsilon_d$ from ph symmetry.
    The spectra have been normalized to their maximum value in the considered energy window and shifted for
    better visibility. The inset shows the deviation of the occupancy from ph symmetry $\langle\delta{N}_d\rangle=\langle{N_d}\rangle-N=\alpha$
    as a function of of $\delta\epsilon$.
  }
\end{figure}

Finally, let us revisit the case of an $S=3/2$ quantum magnet with positive uniaxial anisotropy $D>0$
coupled to a conduction electron bath. This models the situation of Co adatoms deposited on large CuN$_2$
islands for which the renormalization of single-ion magnetic anisotropy by Kondo exchange was described
first~\cite{Oberg:NatNano:2014}. This can be modelled by a three-orbital Anderson model at and around
half-filling, so that the Hund's rule coupling leads to a spin-3/2 GS for the isolated impurity.
The MA term in (\ref{eq:imp}) with $D>0$ then splits the GS quadruplet
into two doublets as illustrated in Fig.~\ref{fig:model}(c). Assuming the same interaction parameters
as for the spin-1 case above ($U=3.5$eV, $U^\prime=2.5$eV and $J_H=0.5$eV), in order to achieve exactly
the half-filled (i.e. ph symmetric) case, the energies $\epsilon_d$ of the three degenerate impurity levels
have to be tuned to $\epsilon_d^\ast=-6.25$eV according to (\ref{eq:ed_ph}) with the number of electrons $N=3$ now.
Also in the case of the spin-3/2 quantum magnet valence fluctuations induced by detuning of the impurity levels
from ph symmetry lead to the renormalization of the spin excitation energies, as can be seen in Fig.~\ref{fig:spin3/2}.
At the same time the Kondo peak grows and becomes increasingly asymmetric.
Generally the spectra become more asymmetric due to the lifting of the ph symmetry, although
the asymmetric character is somewhat less pronounced than in the case of the spin-1 quantum magnet.
As can be seen from the inset of Fig.~\ref{fig:spin3/2}, the induced \emph{total} charge fluctuations, measured
by the deviation of the total impurity shell occupancy from ph symmetry, $\langle\delta{N}_d\rangle=\langle{N_d}\rangle-N=\alpha$,
are of the same order as for the spin-1 case (Fig.~\ref{fig:spin1_Ed}). 
However, due to the larger number of orbitals for the spin-3/2 model, the charge fluctuations \emph{per orbital} are smaller by a factor of 2/3,
explaining the somewhat weaker breaking of ph symmetry here. 

For Co atoms on large CuN islands, in addition to the already described Kondo exchange coupling to the 
conduction electrons also charge fluctuations might contribute to the observed shift of the spin excitation steps to 
lower energies and the concomitant growth of the Kondo peak, when the Co adsorption site moves closer to the border 
of the island.\cite{Oberg:NatNano:2014}

\section{Conclusions}
\label{sec:conclusions}

To conclude, the Kondo exchange coupling of a quantum spin to the conduction electrons in the substrate
leads to the renormalization of the spin excitation energies associated with MA, even in the complete 
absence of the Kondo effect. This is relevant for example, for integer spin quantum magnets with positive
uniaxial MA where the Kondo effect cannot take place due to the lack of GS degeneracy. This effect has
recently been observed for an FePc molecule deposited on a Au substrate:\cite{Hiraoka:NatComm:2017}
the Kondo peak at zero bias, observed in the $dI/dV$ in tunneling regime, develops into two step features 
situated symmetrically around zero at positive and negative finite bias, typical for the spin excitations 
associated with the MA of a spin-1 quantum magnet, when the molecule is lifted from the substrate using the 
STM tip, thus reducing the Kondo exchange coupling of the spin carrying orbitals of the Fe center with the
conduction electrons. These findings highlight once more\cite{Oberg:NatNano:2014,Delgado:SS:2014,Hiraoka:NatComm:2017}
that inelastic spin flip excitations and Kondo effect are really two sides of the same coin.

A second important finding is that similar to the renormalization by Kondo exchange, also charge fluctuations,
induced by detuning of the impurity shell away from integer occupancy, lead to the renormalization of the spin
excitation energies, both in the absence and in the presence of the Kondo effect. In the regime of very strong
charge fluctuations, the combined effect of breaking the ph symmetry plus the spin excitation steps moving to
lower energies may lead to a Fano-like feature in the spectral function and corresponding $dI/dV$ spectra, for a
spin-1 quantum magnet with positive uniaxial anisotropy i.e. in the complete absence of the Kondo effect.
Hence Fano-like lineshapes, which are usually taken as evidence for the Kondo effect in STM spectroscopy
of magnetic adatoms and molecules on conducting substrates, should always be taken with a grain of salt. 
With respect to the case of Co on CuN islands,\cite{Oberg:NatNano:2014} which can be modeled by spin-3/2 impurities subject to positive
uniaxial anisotropy, also renormalization by charge fluctuations instead of or in addition to the
proposed mechanism of renormalization by Kondo exchange could explain the observed reduction of spin excitation
energies (and concomitant growth of the Kondo peak) for Co adsorption sites closer to the border of the CuN island.

\begin{acknowledgments}
  I am grateful to J. Fern\'andez-Rossier and N. Lorente for fruitful discussions.
  Financial support by “Grupos Consolidados UPV/EHU del Gobierno Vasco” (IT578-13) is gratefully acknowledged.
\end{acknowledgments}

%%%%%%%%%%%%%%%%%%%%%%%%%%%%%%%%%%%%%%%%%%%%%%%%%%%%%%%%%%%%%%%%%%%%%%%%%%%%%%%%%%%%%%%%%%
\appendix
%%%%%%%%%%%%%%%%%%%%%%%%%%%%%%%%%%%%%%%%%%%%%%%%%%%%%%%%%%%%%%%%%%%%%%%%%%%%%%%%%%%%%%%%%%

\section{Comparison with NCA for spin-1 quantum magnet with positive uniaxial MA}
\label{app:nca_artifact}

\begin{figure}[h]
  \begin{center}
    \includegraphics[width=0.9\linewidth]{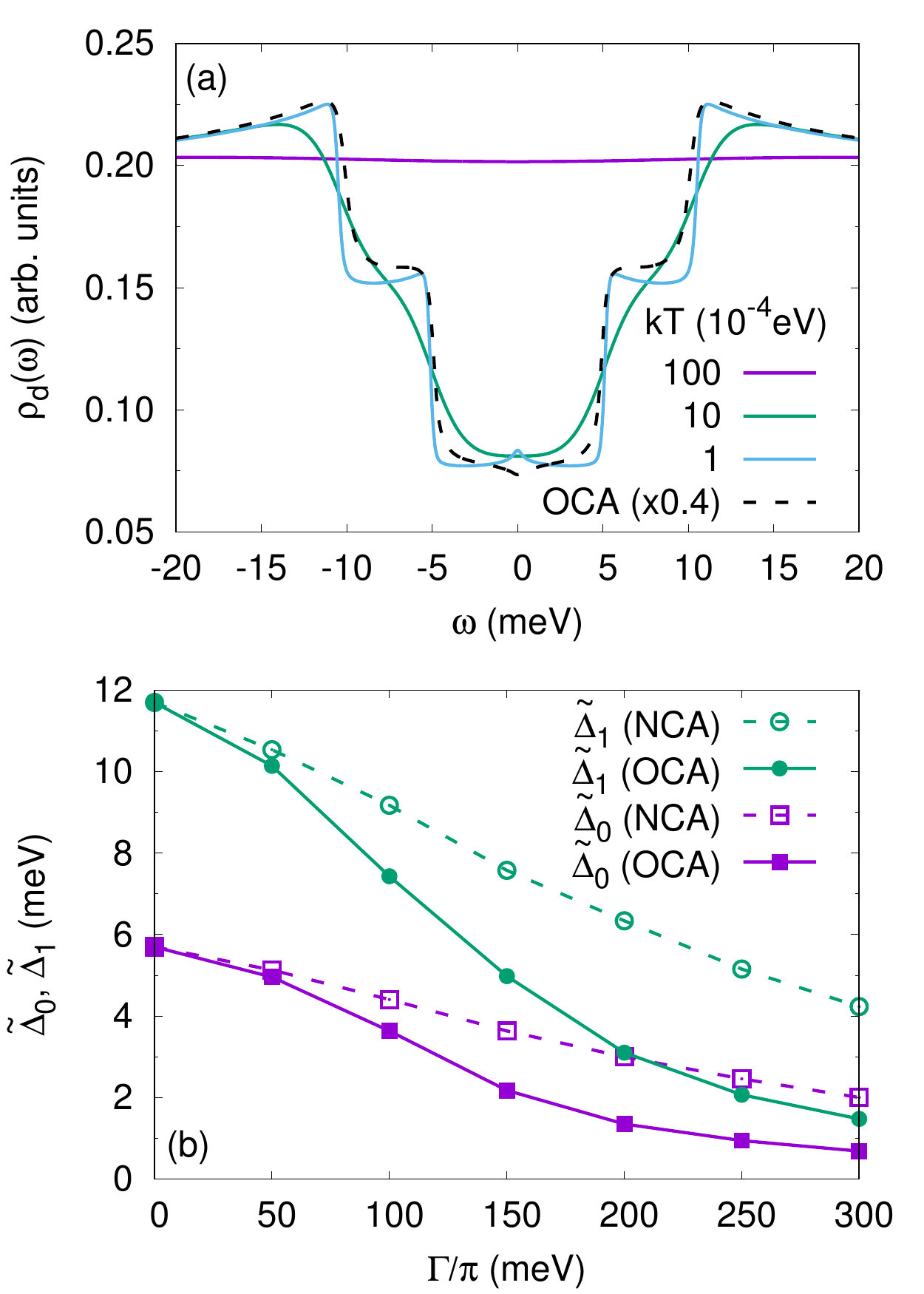}
  \end{center}
  \caption{
    \label{fig:spin1_nca}
    (a) NCA spectra for an $S=1$ quantum magnet with same parameters as in Fig.~\ref{fig:spin1_temp}(b):
    $D=8.7$meV, $E=3$meV, $\Gamma/\pi=50$meV, $U=3.5$eV, $U^\prime=2.5$eV, $J_{\rm H}=0.5$eV, and
    $\epsilon_d=-4$eV (ph symmetric case). For comparison also the corresponding OCA spectrum (scaled by a factor
    of 0.4) is shown for the lowest temperature ($kT=10^{-4}$eV).
    (b) Comparison of effective spin excitation energies  $\widetilde\Delta_0$ and $\widetilde\Delta_1$
    calculated by NCA (dashed lines, open symbols) and by OCA (full lines and full symbols) as a function
    of the hybridization $\Gamma$ for same parameters as in Fig.~\ref{fig:spin1_hyb}.
  }
\end{figure}

Fig.~\ref{fig:spin1_nca}(a) shows the spectra of a spin-1 quantum magnet for the same
parameters as those calculated by OCA shown in Fig.~\ref{fig:spin1_temp}(b) for different temperatures,
but now calculated within the simpler Non-Crossing Approximation (NCA), only including diagrams
where conduction elctron lines are not allowed to cross. For comparison also the corresponding
OCA spectrum for the lowest temperature is shown. Note that within OCA the spin flip steps are higher
by a factor $\sim$2.5 compared to NCA. Thus for better visibility the OCA spectrum has been scaled by
a factor of 0.4.

Overall the spectra and their temperature evolution are quite similar for both approximations.
However, NCA gives rise to a Kondo-like peak artifact in the spectrum at low temperatures, which is absent
in the spectrum obtained within the more sophisticated OCA. Apparently the vertex corrections over the NCA
bubble diagram present in the OCA by taking into account the crossing diagrams %%(\ref{eq:pp_self_oca})
remedy this problem of the NCA. On the other hand a small dip-like feature appears around the Fermi level
for low temperatures in the OCA spectrum, as can be seen in Fig.~\ref{fig:spin1_temp}(b), which might be an
artifact of that approximation. Also note that the ``up-bending'' of the inner spin flip steps at low temperatures
is much less pronounced within OCA than within NCA, but comparable for the outer spin flip steps.
On the other hand the renormalization of the spin flip steps is somewhat stronger within OCA.

As can be seen from Fig.~\ref{fig:spin1_nca}(b) which compares the effective spin flip excitations
calculated by NCA and OCA as a function of the hybridization $\Gamma$, NCA generally underestimates the renormalization
of spin flip excitations as compared to OCA. We can also see that although NCA also predicts a red shift
of the spin flip excitations due to Kondo exchange coupling, the behavior of the renormalization
as a function of the hybridization is qualitatively different for both levels of approximation: While
the red shift of the spin flip excitation energies $\widetilde\Delta_0$ and $\widetilde\Delta_1$
as a function of $\Gamma$ follows a linear behavior within NCA, it is of approximately quadratic form within
OCA for small to intermediate hybridization strengths, becoming asymptotically constant for larger values of $\Gamma$. 
Hence overall NCA is capable of qualitatively reproducing the OCA spectra (with the exception of the
Kondo artifact), but disagrees on the quantitative level with OCA.
As one might expect for approximations based on the expansion around the atomic limit,
the quantitative disagreement becomes worse with increasing hybridization strength.

%%%%%%%%%%%%%%%%%%%%%%%%%%%%%%%%%%%%%%%%%%%%%%%%%%%%%%%%%%%%%%%%%%%%%%%%%%%%%%%%%%%%%%%%%%%%%%%%%%%%%%%%%%%%%%%%%%%%%%%%%%%%%%%%%%%%%%%%%%%%%%%%%%%

\bibliography{nanodmft}

\begin{thebibliography}{47}%
\makeatletter
\providecommand \@ifxundefined [1]{%
 \@ifx{#1\undefined}
}%
\providecommand \@ifnum [1]{%
 \ifnum #1\expandafter \@firstoftwo
 \else \expandafter \@secondoftwo
 \fi
}%
\providecommand \@ifx [1]{%
 \ifx #1\expandafter \@firstoftwo
 \else \expandafter \@secondoftwo
 \fi
}%
\providecommand \natexlab [1]{#1}%
\providecommand \enquote  [1]{``#1''}%
\providecommand \bibnamefont  [1]{#1}%
\providecommand \bibfnamefont [1]{#1}%
\providecommand \citenamefont [1]{#1}%
\providecommand \href@noop [0]{\@secondoftwo}%
\providecommand \href [0]{\begingroup \@sanitize@url \@href}%
\providecommand \@href[1]{\@@startlink{#1}\@@href}%
\providecommand \@@href[1]{\endgroup#1\@@endlink}%
\providecommand \@sanitize@url [0]{\catcode `\\12\catcode `\$12\catcode
  `\&12\catcode `\#12\catcode `\^12\catcode `\_12\catcode `\%12\relax}%
\providecommand \@@startlink[1]{}%
\providecommand \@@endlink[0]{}%
\providecommand \url  [0]{\begingroup\@sanitize@url \@url }%
\providecommand \@url [1]{\endgroup\@href {#1}{\urlprefix }}%
\providecommand \urlprefix  [0]{URL }%
\providecommand \Eprint [0]{\href }%
\providecommand \doibase [0]{http://dx.doi.org/}%
\providecommand \selectlanguage [0]{\@gobble}%
\providecommand \bibinfo  [0]{\@secondoftwo}%
\providecommand \bibfield  [0]{\@secondoftwo}%
\providecommand \translation [1]{[#1]}%
\providecommand \BibitemOpen [0]{}%
\providecommand \bibitemStop [0]{}%
\providecommand \bibitemNoStop [0]{.\EOS\space}%
\providecommand \EOS [0]{\spacefactor3000\relax}%
\providecommand \BibitemShut  [1]{\csname bibitem#1\endcsname}%
\let\auto@bib@innerbib\@empty
%</preamble>
\bibitem [{\citenamefont {Hewson}(1997)}]{Hewson:book:1997}%
  \BibitemOpen
  \bibfield  {author} {\bibinfo {author} {\bibfnamefont {A.~C.}\ \bibnamefont
  {Hewson}},\ }\href@noop {} {\emph {\bibinfo {title} {The Kondo problem to
  heavy fermions}}}\ (\bibinfo  {publisher} {Cambr. Univ. Press},\ \bibinfo
  {address} {Cambridge},\ \bibinfo {year} {1997})\BibitemShut {NoStop}%
\bibitem [{\citenamefont {Gatteschi}\ \emph {et~al.}(2006)\citenamefont
  {Gatteschi}, \citenamefont {Sessoli},\ and\ \citenamefont
  {Villain}}]{Gatteschi:book:2006}%
  \BibitemOpen
  \bibfield  {author} {\bibinfo {author} {\bibfnamefont {D.}~\bibnamefont
  {Gatteschi}}, \bibinfo {author} {\bibfnamefont {R.}~\bibnamefont {Sessoli}},
  \ and\ \bibinfo {author} {\bibfnamefont {J.}~\bibnamefont {Villain}},\
  }\href@noop {} {\emph {\bibinfo {title} {{Molecular Nanomagnets}}}}\
  (\bibinfo  {publisher} {Oxford University Press},\ \bibinfo {address}
  {Oxford},\ \bibinfo {year} {2006})\BibitemShut {NoStop}%
\bibitem [{\citenamefont {Madhavan}\ \emph {et~al.}(1998)\citenamefont
  {Madhavan}, \citenamefont {Chen}, \citenamefont {Jamneala}, \citenamefont
  {Crommie},\ and\ \citenamefont {Wingreen}}]{Madhavan:Science:1998}%
  \BibitemOpen
  \bibfield  {author} {\bibinfo {author} {\bibfnamefont {V.}~\bibnamefont
  {Madhavan}}, \bibinfo {author} {\bibfnamefont {W.}~\bibnamefont {Chen}},
  \bibinfo {author} {\bibfnamefont {T.}~\bibnamefont {Jamneala}}, \bibinfo
  {author} {\bibfnamefont {M.~F.}\ \bibnamefont {Crommie}}, \ and\ \bibinfo
  {author} {\bibfnamefont {N.~S.}\ \bibnamefont {Wingreen}},\ }\href {\doibase
  10.1126/science.280.5363.567} {\bibfield  {journal} {\bibinfo  {journal}
  {Science}\ }\textbf {\bibinfo {volume} {280}},\ \bibinfo {pages} {567}
  (\bibinfo {year} {1998})}\BibitemShut {NoStop}%
\bibitem [{\citenamefont {Li}\ \emph {et~al.}(1998)\citenamefont {Li},
  \citenamefont {Schneider}, \citenamefont {Berndt},\ and\ \citenamefont
  {Delley}}]{Li:PRL:1998}%
  \BibitemOpen
  \bibfield  {author} {\bibinfo {author} {\bibfnamefont {J.}~\bibnamefont
  {Li}}, \bibinfo {author} {\bibfnamefont {W.-D.}\ \bibnamefont {Schneider}},
  \bibinfo {author} {\bibfnamefont {R.}~\bibnamefont {Berndt}}, \ and\ \bibinfo
  {author} {\bibfnamefont {B.}~\bibnamefont {Delley}},\ }\href {\doibase
  10.1103/PhysRevLett.80.2893} {\bibfield  {journal} {\bibinfo  {journal}
  {Phys. Rev. Lett.}\ }\textbf {\bibinfo {volume} {80}},\ \bibinfo {pages}
  {2893} (\bibinfo {year} {1998})}\BibitemShut {NoStop}%
\bibitem [{\citenamefont {Zhao}\ \emph {et~al.}(2005)\citenamefont {Zhao},
  \citenamefont {Li}, \citenamefont {Chen}, \citenamefont {Xiang},
  \citenamefont {Wang}, \citenamefont {Pan}, \citenamefont {Wang},
  \citenamefont {Xiao}, \citenamefont {Yang}, \citenamefont {Hou},\ and\
  \citenamefont {Zhu}}]{Zhao:Science:2005}%
  \BibitemOpen
  \bibfield  {author} {\bibinfo {author} {\bibfnamefont {A.}~\bibnamefont
  {Zhao}}, \bibinfo {author} {\bibfnamefont {Q.}~\bibnamefont {Li}}, \bibinfo
  {author} {\bibfnamefont {L.}~\bibnamefont {Chen}}, \bibinfo {author}
  {\bibfnamefont {H.}~\bibnamefont {Xiang}}, \bibinfo {author} {\bibfnamefont
  {W.}~\bibnamefont {Wang}}, \bibinfo {author} {\bibfnamefont {S.}~\bibnamefont
  {Pan}}, \bibinfo {author} {\bibfnamefont {B.}~\bibnamefont {Wang}}, \bibinfo
  {author} {\bibfnamefont {X.}~\bibnamefont {Xiao}}, \bibinfo {author}
  {\bibfnamefont {J.}~\bibnamefont {Yang}}, \bibinfo {author} {\bibfnamefont
  {J.~G.}\ \bibnamefont {Hou}}, \ and\ \bibinfo {author} {\bibfnamefont
  {Q.}~\bibnamefont {Zhu}},\ }\href {\doibase 10.1126/science.1113449}
  {\bibfield  {journal} {\bibinfo  {journal} {Science}\ }\textbf {\bibinfo
  {volume} {309}},\ \bibinfo {pages} {1542} (\bibinfo {year}
  {2005})}\BibitemShut {NoStop}%
\bibitem [{\citenamefont {Iancu}\ \emph {et~al.}(2006)\citenamefont {Iancu},
  \citenamefont {Deshpande},\ and\ \citenamefont {Hla}}]{Iancu:NL:2006}%
  \BibitemOpen
  \bibfield  {author} {\bibinfo {author} {\bibfnamefont {V.}~\bibnamefont
  {Iancu}}, \bibinfo {author} {\bibfnamefont {A.}~\bibnamefont {Deshpande}}, \
  and\ \bibinfo {author} {\bibfnamefont {S.-W.}\ \bibnamefont {Hla}},\ }\href
  {\doibase 10.1021/nl0601886} {\bibfield  {journal} {\bibinfo  {journal} {Nano
  Letters}\ }\textbf {\bibinfo {volume} {6}},\ \bibinfo {pages} {820} (\bibinfo
  {year} {2006})}\BibitemShut {NoStop}%
\bibitem [{\citenamefont {Hirjibehedin}\ \emph {et~al.}(2007)\citenamefont
  {Hirjibehedin}, \citenamefont {Lin}, \citenamefont {Otte}, \citenamefont
  {Ternes}, \citenamefont {Lutz}, \citenamefont {Jones},\ and\ \citenamefont
  {Heinrich}}]{Hirjibehedin:Science:2007}%
  \BibitemOpen
  \bibfield  {author} {\bibinfo {author} {\bibfnamefont {C.~F.}\ \bibnamefont
  {Hirjibehedin}}, \bibinfo {author} {\bibfnamefont {C.-Y.}\ \bibnamefont
  {Lin}}, \bibinfo {author} {\bibfnamefont {A.~F.}\ \bibnamefont {Otte}},
  \bibinfo {author} {\bibfnamefont {M.}~\bibnamefont {Ternes}}, \bibinfo
  {author} {\bibfnamefont {C.~P.}\ \bibnamefont {Lutz}}, \bibinfo {author}
  {\bibfnamefont {B.~A.}\ \bibnamefont {Jones}}, \ and\ \bibinfo {author}
  {\bibfnamefont {A.~J.}\ \bibnamefont {Heinrich}},\ }\href {\doibase
  10.1126/science.1146110} {\bibfield  {journal} {\bibinfo  {journal}
  {Science}\ }\textbf {\bibinfo {volume} {317}},\ \bibinfo {pages}
  {1199–1203} (\bibinfo {year} {2007})}\BibitemShut {NoStop}%
\bibitem [{\citenamefont {Yu}\ \emph {et~al.}(2005)\citenamefont {Yu},
  \citenamefont {Keane}, \citenamefont {Ciszek}, \citenamefont {Cheng},
  \citenamefont {Tour}, \citenamefont {Baruah}, \citenamefont {Pederson},\ and\
  \citenamefont {Natelson}}]{Yu:PRL:2005}%
  \BibitemOpen
  \bibfield  {author} {\bibinfo {author} {\bibfnamefont {L.~H.}\ \bibnamefont
  {Yu}}, \bibinfo {author} {\bibfnamefont {Z.~K.}\ \bibnamefont {Keane}},
  \bibinfo {author} {\bibfnamefont {J.~W.}\ \bibnamefont {Ciszek}}, \bibinfo
  {author} {\bibfnamefont {L.}~\bibnamefont {Cheng}}, \bibinfo {author}
  {\bibfnamefont {J.~M.}\ \bibnamefont {Tour}}, \bibinfo {author}
  {\bibfnamefont {T.}~\bibnamefont {Baruah}}, \bibinfo {author} {\bibfnamefont
  {M.~R.}\ \bibnamefont {Pederson}}, \ and\ \bibinfo {author} {\bibfnamefont
  {D.}~\bibnamefont {Natelson}},\ }\href {\doibase
  10.1103/PhysRevLett.95.256803} {\bibfield  {journal} {\bibinfo  {journal}
  {Phys. Rev. Lett.}\ }\textbf {\bibinfo {volume} {95}},\ \bibinfo {pages}
  {256803} (\bibinfo {year} {2005})}\BibitemShut {NoStop}%
\bibitem [{\citenamefont {Parks}\ \emph {et~al.}(2007)\citenamefont {Parks},
  \citenamefont {Champagne}, \citenamefont {Hutchison}, \citenamefont
  {Flores-Torres}, \citenamefont {Abru\~na},\ and\ \citenamefont
  {Ralph}}]{Parks:PRL:2007}%
  \BibitemOpen
  \bibfield  {author} {\bibinfo {author} {\bibfnamefont {J.~J.}\ \bibnamefont
  {Parks}}, \bibinfo {author} {\bibfnamefont {A.~R.}\ \bibnamefont
  {Champagne}}, \bibinfo {author} {\bibfnamefont {G.~R.}\ \bibnamefont
  {Hutchison}}, \bibinfo {author} {\bibfnamefont {S.}~\bibnamefont
  {Flores-Torres}}, \bibinfo {author} {\bibfnamefont {H.~D.}\ \bibnamefont
  {Abru\~na}}, \ and\ \bibinfo {author} {\bibfnamefont {D.~C.}\ \bibnamefont
  {Ralph}},\ }\href {\doibase 10.1103/PhysRevLett.99.026601} {\bibfield
  {journal} {\bibinfo  {journal} {Phys. Rev. Lett.}\ }\textbf {\bibinfo
  {volume} {99}},\ \bibinfo {pages} {026601} (\bibinfo {year}
  {2007})}\BibitemShut {NoStop}%
\bibitem [{\citenamefont {Calvo}\ \emph {et~al.}(2009)\citenamefont {Calvo},
  \citenamefont {Fern\'andez-Rossier}, \citenamefont {Palacios}, \citenamefont
  {Jacob}, \citenamefont {Natelson},\ and\ \citenamefont
  {Untiedt}}]{Calvo:Nature:2009}%
  \BibitemOpen
  \bibfield  {author} {\bibinfo {author} {\bibfnamefont {M.~R.}\ \bibnamefont
  {Calvo}}, \bibinfo {author} {\bibfnamefont {J.}~\bibnamefont
  {Fern\'andez-Rossier}}, \bibinfo {author} {\bibfnamefont {J.~J.}\
  \bibnamefont {Palacios}}, \bibinfo {author} {\bibfnamefont {D.}~\bibnamefont
  {Jacob}}, \bibinfo {author} {\bibfnamefont {D.}~\bibnamefont {Natelson}}, \
  and\ \bibinfo {author} {\bibfnamefont {C.}~\bibnamefont {Untiedt}},\ }\href
  {\doibase 10.1038/nature07878} {\bibfield  {journal} {\bibinfo  {journal}
  {Nature}\ }\textbf {\bibinfo {volume} {358}},\ \bibinfo {pages} {1150}
  (\bibinfo {year} {2009})}\BibitemShut {NoStop}%
\bibitem [{\citenamefont {Parks}\ \emph {et~al.}(2010)\citenamefont {Parks},
  \citenamefont {Champagne}, \citenamefont {Costi}, \citenamefont {Shum},
  \citenamefont {Pasupathy}, \citenamefont {Neuscamman}, \citenamefont
  {Flores-Torres}, \citenamefont {Cornaglia}, \citenamefont {Aligia},
  \citenamefont {Balseiro}, \citenamefont {Chan}, \citenamefont {Abru\~na},\
  and\ \citenamefont {Ralph}}]{Parks:Science:2010}%
  \BibitemOpen
  \bibfield  {author} {\bibinfo {author} {\bibfnamefont {J.~J.}\ \bibnamefont
  {Parks}}, \bibinfo {author} {\bibfnamefont {A.~R.}\ \bibnamefont
  {Champagne}}, \bibinfo {author} {\bibfnamefont {T.~A.}\ \bibnamefont
  {Costi}}, \bibinfo {author} {\bibfnamefont {W.~W.}\ \bibnamefont {Shum}},
  \bibinfo {author} {\bibfnamefont {A.~N.}\ \bibnamefont {Pasupathy}}, \bibinfo
  {author} {\bibfnamefont {E.}~\bibnamefont {Neuscamman}}, \bibinfo {author}
  {\bibfnamefont {S.}~\bibnamefont {Flores-Torres}}, \bibinfo {author}
  {\bibfnamefont {P.~S.}\ \bibnamefont {Cornaglia}}, \bibinfo {author}
  {\bibfnamefont {A.~A.}\ \bibnamefont {Aligia}}, \bibinfo {author}
  {\bibfnamefont {C.~A.}\ \bibnamefont {Balseiro}}, \bibinfo {author}
  {\bibfnamefont {G.~K.-L.}\ \bibnamefont {Chan}}, \bibinfo {author}
  {\bibfnamefont {H.~D.}\ \bibnamefont {Abru\~na}}, \ and\ \bibinfo {author}
  {\bibfnamefont {D.~C.}\ \bibnamefont {Ralph}},\ }\href {\doibase
  10.1126/science.1186874} {\bibfield  {journal} {\bibinfo  {journal}
  {Science}\ }\textbf {\bibinfo {volume} {328}},\ \bibinfo {pages} {1370}
  (\bibinfo {year} {2010})}\BibitemShut {NoStop}%
\bibitem [{\citenamefont {Schiller}\ and\ \citenamefont
  {Hershfield}(2000)}]{Schiller:PRB:2000}%
  \BibitemOpen
  \bibfield  {author} {\bibinfo {author} {\bibfnamefont {A.}~\bibnamefont
  {Schiller}}\ and\ \bibinfo {author} {\bibfnamefont {S.}~\bibnamefont
  {Hershfield}},\ }\href {\doibase 10.1103/PhysRevB.61.9036} {\bibfield
  {journal} {\bibinfo  {journal} {Phys. Rev. B}\ }\textbf {\bibinfo {volume}
  {61}},\ \bibinfo {pages} {9036} (\bibinfo {year} {2000})}\BibitemShut
  {NoStop}%
\bibitem [{\citenamefont {\'Ujs\'aghy}\ \emph {et~al.}(2000)\citenamefont
  {\'Ujs\'aghy}, \citenamefont {Kroha}, \citenamefont {Szunyogh},\ and\
  \citenamefont {Zawadowski}}]{Ujsaghy:PRL:2000}%
  \BibitemOpen
  \bibfield  {author} {\bibinfo {author} {\bibfnamefont {O.}~\bibnamefont
  {\'Ujs\'aghy}}, \bibinfo {author} {\bibfnamefont {J.}~\bibnamefont {Kroha}},
  \bibinfo {author} {\bibfnamefont {L.}~\bibnamefont {Szunyogh}}, \ and\
  \bibinfo {author} {\bibfnamefont {A.}~\bibnamefont {Zawadowski}},\ }\href
  {\doibase 10.1103/PhysRevLett.85.2557} {\bibfield  {journal} {\bibinfo
  {journal} {Phys. Rev. Lett.}\ }\textbf {\bibinfo {volume} {85}},\ \bibinfo
  {pages} {2557} (\bibinfo {year} {2000})}\BibitemShut {NoStop}%
\bibitem [{\citenamefont
  {Fern\'andez-Rossier}(2009)}]{Fernandez-Rossier:PRL:2009}%
  \BibitemOpen
  \bibfield  {author} {\bibinfo {author} {\bibfnamefont {J.}~\bibnamefont
  {Fern\'andez-Rossier}},\ }\href {\doibase 10.1103/PhysRevLett.102.256802}
  {\bibfield  {journal} {\bibinfo  {journal} {Phys. Rev. Lett.}\ }\textbf
  {\bibinfo {volume} {102}},\ \bibinfo {pages} {256802} (\bibinfo {year}
  {2009})}\BibitemShut {NoStop}%
\bibitem [{\citenamefont {Otte}\ \emph {et~al.}(2008)\citenamefont {Otte},
  \citenamefont {Ternes}, \citenamefont {von Bergmann}, \citenamefont {Loth},
  \citenamefont {Brune}, \citenamefont {Lutz}, \citenamefont {Hirjibehedin},\
  and\ \citenamefont {Heinrich}}]{Otte:NatPhys:2008}%
  \BibitemOpen
  \bibfield  {author} {\bibinfo {author} {\bibfnamefont {A.~F.}\ \bibnamefont
  {Otte}}, \bibinfo {author} {\bibfnamefont {M.}~\bibnamefont {Ternes}},
  \bibinfo {author} {\bibfnamefont {K.}~\bibnamefont {von Bergmann}}, \bibinfo
  {author} {\bibfnamefont {S.}~\bibnamefont {Loth}}, \bibinfo {author}
  {\bibfnamefont {H.}~\bibnamefont {Brune}}, \bibinfo {author} {\bibfnamefont
  {C.~P.}\ \bibnamefont {Lutz}}, \bibinfo {author} {\bibfnamefont {C.~F.}\
  \bibnamefont {Hirjibehedin}}, \ and\ \bibinfo {author} {\bibfnamefont
  {A.~J.}\ \bibnamefont {Heinrich}},\ }\href {\doibase 10.1038/nphys1072}
  {\bibfield  {journal} {\bibinfo  {journal} {Nature Physics}\ }\textbf
  {\bibinfo {volume} {4}},\ \bibinfo {pages} {847} (\bibinfo {year}
  {2008})}\BibitemShut {NoStop}%
\bibitem [{\citenamefont {Tsukahara}\ \emph {et~al.}(2011)\citenamefont
  {Tsukahara}, \citenamefont {Shiraki}, \citenamefont {Itou}, \citenamefont
  {Ohta}, \citenamefont {Takagi},\ and\ \citenamefont
  {Kawai}}]{Tsukahara:PRL:2011}%
  \BibitemOpen
  \bibfield  {author} {\bibinfo {author} {\bibfnamefont {N.}~\bibnamefont
  {Tsukahara}}, \bibinfo {author} {\bibfnamefont {S.}~\bibnamefont {Shiraki}},
  \bibinfo {author} {\bibfnamefont {S.}~\bibnamefont {Itou}}, \bibinfo {author}
  {\bibfnamefont {N.}~\bibnamefont {Ohta}}, \bibinfo {author} {\bibfnamefont
  {N.}~\bibnamefont {Takagi}}, \ and\ \bibinfo {author} {\bibfnamefont
  {M.}~\bibnamefont {Kawai}},\ }\href {\doibase 10.1103/PhysRevLett.106.187201}
  {\bibfield  {journal} {\bibinfo  {journal} {Phys. Rev. Lett.}\ }\textbf
  {\bibinfo {volume} {106}},\ \bibinfo {pages} {187201} (\bibinfo {year}
  {2011})}\BibitemShut {NoStop}%
\bibitem [{\citenamefont {Tsukahara}\ \emph {et~al.}(2009)\citenamefont
  {Tsukahara}, \citenamefont {Noto}, \citenamefont {Ohara}, \citenamefont
  {Shiraki}, \citenamefont {Takagi}, \citenamefont {Takata}, \citenamefont
  {Miyawaki}, \citenamefont {Taguchi}, \citenamefont {Chainani}, \citenamefont
  {Shin},\ and\ \citenamefont {Kawai}}]{Tsukahara:PRL:2009}%
  \BibitemOpen
  \bibfield  {author} {\bibinfo {author} {\bibfnamefont {N.}~\bibnamefont
  {Tsukahara}}, \bibinfo {author} {\bibfnamefont {K.~I.}\ \bibnamefont {Noto}},
  \bibinfo {author} {\bibfnamefont {M.}~\bibnamefont {Ohara}}, \bibinfo
  {author} {\bibfnamefont {S.}~\bibnamefont {Shiraki}}, \bibinfo {author}
  {\bibfnamefont {N.}~\bibnamefont {Takagi}}, \bibinfo {author} {\bibfnamefont
  {Y.}~\bibnamefont {Takata}}, \bibinfo {author} {\bibfnamefont
  {J.}~\bibnamefont {Miyawaki}}, \bibinfo {author} {\bibfnamefont
  {M.}~\bibnamefont {Taguchi}}, \bibinfo {author} {\bibfnamefont
  {A.}~\bibnamefont {Chainani}}, \bibinfo {author} {\bibfnamefont
  {S.}~\bibnamefont {Shin}}, \ and\ \bibinfo {author} {\bibfnamefont
  {M.}~\bibnamefont {Kawai}},\ }\href {\doibase 10.1103/PhysRevLett.102.167203}
  {\bibfield  {journal} {\bibinfo  {journal} {Phys. Rev. Lett.}\ }\textbf
  {\bibinfo {volume} {102}},\ \bibinfo {pages} {167203} (\bibinfo {year}
  {2009})}\BibitemShut {NoStop}%
\bibitem [{\citenamefont {Oberg}\ \emph {et~al.}(2014)\citenamefont {Oberg},
  \citenamefont {Calvo}, \citenamefont {Delgado}, \citenamefont {Moro-Lagares},
  \citenamefont {Serrate}, \citenamefont {Jacob}, \citenamefont
  {Fernandez-Rossier},\ and\ \citenamefont
  {Hirjibehedin}}]{Oberg:NatNano:2014}%
  \BibitemOpen
  \bibfield  {author} {\bibinfo {author} {\bibfnamefont {J.~C.}\ \bibnamefont
  {Oberg}}, \bibinfo {author} {\bibfnamefont {M.~R.}\ \bibnamefont {Calvo}},
  \bibinfo {author} {\bibfnamefont {F.}~\bibnamefont {Delgado}}, \bibinfo
  {author} {\bibfnamefont {M.}~\bibnamefont {Moro-Lagares}}, \bibinfo {author}
  {\bibfnamefont {D.}~\bibnamefont {Serrate}}, \bibinfo {author} {\bibfnamefont
  {D.}~\bibnamefont {Jacob}}, \bibinfo {author} {\bibfnamefont
  {J.}~\bibnamefont {Fernandez-Rossier}}, \ and\ \bibinfo {author}
  {\bibfnamefont {C.~F.}\ \bibnamefont {Hirjibehedin}},\ }\href {\doibase
  10.1038/nnano.2013.264} {\bibfield  {journal} {\bibinfo  {journal} {Nature
  Nano}\ }\textbf {\bibinfo {volume} {9}},\ \bibinfo {pages} {64} (\bibinfo
  {year} {2014})}\BibitemShut {NoStop}%
\bibitem [{\citenamefont {Hiraoka}\ \emph {et~al.}(2017)\citenamefont
  {Hiraoka}, \citenamefont {Minamitani}, \citenamefont {Arafune}, \citenamefont
  {Tsukahara}, \citenamefont {Watanabe}, \citenamefont {Kawai},\ and\
  \citenamefont {Takagi}}]{Hiraoka:NatComm:2017}%
  \BibitemOpen
  \bibfield  {author} {\bibinfo {author} {\bibfnamefont {R.}~\bibnamefont
  {Hiraoka}}, \bibinfo {author} {\bibfnamefont {E.}~\bibnamefont {Minamitani}},
  \bibinfo {author} {\bibfnamefont {R.}~\bibnamefont {Arafune}}, \bibinfo
  {author} {\bibfnamefont {N.}~\bibnamefont {Tsukahara}}, \bibinfo {author}
  {\bibfnamefont {S.}~\bibnamefont {Watanabe}}, \bibinfo {author}
  {\bibfnamefont {M.}~\bibnamefont {Kawai}}, \ and\ \bibinfo {author}
  {\bibfnamefont {N.}~\bibnamefont {Takagi}},\ }\href {\doibase
  10.1038/ncomms16012} {\bibfield  {journal} {\bibinfo  {journal} {Nature
  Comm.}\ }\textbf {\bibinfo {volume} {8}},\ \bibinfo {pages} {16012} (\bibinfo
  {year} {2017})}\BibitemShut {NoStop}%
\bibitem [{\citenamefont {\v{Z}itko}\ \emph {et~al.}(2008)\citenamefont
  {\v{Z}itko}, \citenamefont {Peters},\ and\ \citenamefont
  {Pruschke}}]{Zitko:PRB:2008}%
  \BibitemOpen
  \bibfield  {author} {\bibinfo {author} {\bibfnamefont {R.}~\bibnamefont
  {\v{Z}itko}}, \bibinfo {author} {\bibfnamefont {R.}~\bibnamefont {Peters}}, \
  and\ \bibinfo {author} {\bibfnamefont {T.}~\bibnamefont {Pruschke}},\ }\href
  {\doibase 10.1103/PhysRevB.78.224404} {\bibfield  {journal} {\bibinfo
  {journal} {Phys. Rev. B}\ }\textbf {\bibinfo {volume} {78}},\ \bibinfo
  {pages} {224404} (\bibinfo {year} {2008})}\BibitemShut {NoStop}%
\bibitem [{\citenamefont {Lorente}\ and\ \citenamefont
  {Gauyacq}(2009)}]{Lorente:PRL:2009}%
  \BibitemOpen
  \bibfield  {author} {\bibinfo {author} {\bibfnamefont {N.}~\bibnamefont
  {Lorente}}\ and\ \bibinfo {author} {\bibfnamefont {J.-P.}\ \bibnamefont
  {Gauyacq}},\ }\href {\doibase 10.1103/PhysRevLett.103.176601} {\bibfield
  {journal} {\bibinfo  {journal} {Phys. Rev. Lett.}\ }\textbf {\bibinfo
  {volume} {103}},\ \bibinfo {pages} {176601} (\bibinfo {year}
  {2009})}\BibitemShut {NoStop}%
\bibitem [{\citenamefont {\v{Z}itko}\ and\ \citenamefont
  {Pruschke}(2010)}]{Zitko:NJP:2010}%
  \BibitemOpen
  \bibfield  {author} {\bibinfo {author} {\bibfnamefont {R.}~\bibnamefont
  {\v{Z}itko}}\ and\ \bibinfo {author} {\bibfnamefont {T.}~\bibnamefont
  {Pruschke}},\ }\href {\doibase 10.1088/1367-2630/12/6/063040} {\bibfield
  {journal} {\bibinfo  {journal} {New. J. Phys.}\ }\textbf {\bibinfo {volume}
  {12}},\ \bibinfo {pages} {063040} (\bibinfo {year} {2010})}\BibitemShut
  {NoStop}%
\bibitem [{\citenamefont {Hurley}\ \emph {et~al.}(2011)\citenamefont {Hurley},
  \citenamefont {Baadji},\ and\ \citenamefont {Sanvito}}]{Hurley:PRB:2011}%
  \BibitemOpen
  \bibfield  {author} {\bibinfo {author} {\bibfnamefont {A.}~\bibnamefont
  {Hurley}}, \bibinfo {author} {\bibfnamefont {N.}~\bibnamefont {Baadji}}, \
  and\ \bibinfo {author} {\bibfnamefont {S.}~\bibnamefont {Sanvito}},\ }\href
  {\doibase 10.1103/PhysRevB.84.115435} {\bibfield  {journal} {\bibinfo
  {journal} {Phys. Rev. B}\ }\textbf {\bibinfo {volume} {84}},\ \bibinfo
  {pages} {115435} (\bibinfo {year} {2011})}\BibitemShut {NoStop}%
\bibitem [{\citenamefont {Delgado}\ \emph {et~al.}(2014)\citenamefont
  {Delgado}, \citenamefont {Hirjibehedin},\ and\ \citenamefont
  {Fern\'andez-Rossier}}]{Delgado:SS:2014}%
  \BibitemOpen
  \bibfield  {author} {\bibinfo {author} {\bibfnamefont {F.}~\bibnamefont
  {Delgado}}, \bibinfo {author} {\bibfnamefont {C.}~\bibnamefont
  {Hirjibehedin}}, \ and\ \bibinfo {author} {\bibfnamefont {J.}~\bibnamefont
  {Fern\'andez-Rossier}},\ }\href {\doibase 10.1016/j.susc.2014.07.009}
  {\bibfield  {journal} {\bibinfo  {journal} {Surf. Sci.}\ }\textbf {\bibinfo
  {volume} {630}},\ \bibinfo {pages} {337} (\bibinfo {year}
  {2014})}\BibitemShut {NoStop}%
\bibitem [{\citenamefont {Ternes}(2015)}]{Ternes:NJP:2015}%
  \BibitemOpen
  \bibfield  {author} {\bibinfo {author} {\bibfnamefont {M.}~\bibnamefont
  {Ternes}},\ }\href {\doibase 10.1088/1367-2630/17/6/063016} {\bibfield
  {journal} {\bibinfo  {journal} {New J. Phys.}\ }\textbf {\bibinfo {volume}
  {17}},\ \bibinfo {pages} {063016} (\bibinfo {year} {2015})}\BibitemShut
  {NoStop}%
\bibitem [{\citenamefont {Ferr\'on}\ \emph {et~al.}(2015)\citenamefont
  {Ferr\'on}, \citenamefont {Lado},\ and\ \citenamefont
  {Fern\'andez-Rossier}}]{Ferron:PRB:2015}%
  \BibitemOpen
  \bibfield  {author} {\bibinfo {author} {\bibfnamefont {A.}~\bibnamefont
  {Ferr\'on}}, \bibinfo {author} {\bibfnamefont {J.}~\bibnamefont {Lado}}, \
  and\ \bibinfo {author} {\bibfnamefont {J.}~\bibnamefont
  {Fern\'andez-Rossier}},\ }\href {\doibase 10.1103/PhysRevB.92.174407}
  {\bibfield  {journal} {\bibinfo  {journal} {Phys. Rev. B}\ }\textbf {\bibinfo
  {volume} {92}},\ \bibinfo {pages} {174407} (\bibinfo {year}
  {2015})}\BibitemShut {NoStop}%
\bibitem [{\citenamefont {Panda}\ \emph {et~al.}(2016)\citenamefont {Panda},
  \citenamefont {Di~Marco}, \citenamefont {Gr\aa{}n\"as}, \citenamefont
  {Eriksson},\ and\ \citenamefont {Fransson}}]{Panda:PRB:2016}%
  \BibitemOpen
  \bibfield  {author} {\bibinfo {author} {\bibfnamefont {S.~K.}\ \bibnamefont
  {Panda}}, \bibinfo {author} {\bibfnamefont {I.}~\bibnamefont {Di~Marco}},
  \bibinfo {author} {\bibfnamefont {O.}~\bibnamefont {Gr\aa{}n\"as}}, \bibinfo
  {author} {\bibfnamefont {O.}~\bibnamefont {Eriksson}}, \ and\ \bibinfo
  {author} {\bibfnamefont {J.}~\bibnamefont {Fransson}},\ }\href {\doibase
  10.1103/PhysRevB.93.140101} {\bibfield  {journal} {\bibinfo  {journal} {Phys.
  Rev. B}\ }\textbf {\bibinfo {volume} {93}},\ \bibinfo {pages} {140101}
  (\bibinfo {year} {2016})}\BibitemShut {NoStop}%
\bibitem [{\citenamefont {Koryt\'ar}\ \emph {et~al.}(2012)\citenamefont
  {Koryt\'ar}, \citenamefont {Lorente},\ and\ \citenamefont
  {Gauyacq}}]{Korytar:PRB:2012}%
  \BibitemOpen
  \bibfield  {author} {\bibinfo {author} {\bibfnamefont {R.}~\bibnamefont
  {Koryt\'ar}}, \bibinfo {author} {\bibfnamefont {N.}~\bibnamefont {Lorente}},
  \ and\ \bibinfo {author} {\bibfnamefont {J.-P.}\ \bibnamefont {Gauyacq}},\
  }\href {\doibase 10.1103/PhysRevB.85.125434} {\bibfield  {journal} {\bibinfo
  {journal} {Phys. Rev. B}\ }\textbf {\bibinfo {volume} {85}},\ \bibinfo
  {pages} {125434} (\bibinfo {year} {2012})}\BibitemShut {NoStop}%
\bibitem [{\citenamefont {Jacobson}\ \emph {et~al.}(2015)\citenamefont
  {Jacobson}, \citenamefont {Herden}, \citenamefont {Muenks}, \citenamefont
  {Laskin}, \citenamefont {Brovko}, \citenamefont {Stepanyuk}, \citenamefont
  {Ternes},\ and\ \citenamefont {Kern}}]{Jacobson:NatComm:2015}%
  \BibitemOpen
  \bibfield  {author} {\bibinfo {author} {\bibfnamefont {P.}~\bibnamefont
  {Jacobson}}, \bibinfo {author} {\bibfnamefont {T.}~\bibnamefont {Herden}},
  \bibinfo {author} {\bibfnamefont {M.}~\bibnamefont {Muenks}}, \bibinfo
  {author} {\bibfnamefont {G.}~\bibnamefont {Laskin}}, \bibinfo {author}
  {\bibfnamefont {O.}~\bibnamefont {Brovko}}, \bibinfo {author} {\bibfnamefont
  {V.}~\bibnamefont {Stepanyuk}}, \bibinfo {author} {\bibfnamefont
  {M.}~\bibnamefont {Ternes}}, \ and\ \bibinfo {author} {\bibfnamefont
  {K.}~\bibnamefont {Kern}},\ }\href {\doibase 10.1038/ncomms9536} {\bibfield
  {journal} {\bibinfo  {journal} {Nature Comm.}\ }\textbf {\bibinfo {volume}
  {6}},\ \bibinfo {pages} {8536} (\bibinfo {year} {2015})}\BibitemShut
  {NoStop}%
\bibitem [{\citenamefont {Jacob}(2015)}]{Jacob:JPCM:2015}%
  \BibitemOpen
  \bibfield  {author} {\bibinfo {author} {\bibfnamefont {D.}~\bibnamefont
  {Jacob}},\ }\href {\doibase 10.1088/0953-8984/27/24/245606} {\bibfield
  {journal} {\bibinfo  {journal} {J. Phys. Condens. Mat.}\ }\textbf {\bibinfo
  {volume} {27}},\ \bibinfo {pages} {245606} (\bibinfo {year}
  {2015})}\BibitemShut {NoStop}%
\bibitem [{\citenamefont {Baruselli}\ \emph {et~al.}(2015)\citenamefont
  {Baruselli}, \citenamefont {Requist}, \citenamefont {Smogunov}, \citenamefont
  {Fabrizio},\ and\ \citenamefont {Tosatti}}]{Baruselli:PRB:2015}%
  \BibitemOpen
  \bibfield  {author} {\bibinfo {author} {\bibfnamefont {P.~P.}\ \bibnamefont
  {Baruselli}}, \bibinfo {author} {\bibfnamefont {R.}~\bibnamefont {Requist}},
  \bibinfo {author} {\bibfnamefont {A.}~\bibnamefont {Smogunov}}, \bibinfo
  {author} {\bibfnamefont {M.}~\bibnamefont {Fabrizio}}, \ and\ \bibinfo
  {author} {\bibfnamefont {E.}~\bibnamefont {Tosatti}},\ }\href {\doibase
  10.1103/PhysRevB.92.045119} {\bibfield  {journal} {\bibinfo  {journal} {Phys.
  Rev. B}\ }\textbf {\bibinfo {volume} {92}},\ \bibinfo {pages} {045119}
  (\bibinfo {year} {2015})}\BibitemShut {NoStop}%
\bibitem [{\citenamefont {Frank}\ and\ \citenamefont
  {Jacob}(2015)}]{Frank:PRB:2015}%
  \BibitemOpen
  \bibfield  {author} {\bibinfo {author} {\bibfnamefont {S.}~\bibnamefont
  {Frank}}\ and\ \bibinfo {author} {\bibfnamefont {D.}~\bibnamefont {Jacob}},\
  }\href {\doibase 10.1103/PhysRevB.92.235127} {\bibfield  {journal} {\bibinfo
  {journal} {Phys. Rev. B}\ }\textbf {\bibinfo {volume} {92}},\ \bibinfo
  {pages} {235127} (\bibinfo {year} {2015})}\BibitemShut {NoStop}%
\bibitem [{\citenamefont {Dang}\ \emph {et~al.}(2016)\citenamefont {Dang},
  \citenamefont {dos Santos~Dias}, \citenamefont {Liebsch},\ and\ \citenamefont
  {Lounis}}]{Dang:PRB:2016}%
  \BibitemOpen
  \bibfield  {author} {\bibinfo {author} {\bibfnamefont {H.~T.}\ \bibnamefont
  {Dang}}, \bibinfo {author} {\bibfnamefont {M.}~\bibnamefont {dos
  Santos~Dias}}, \bibinfo {author} {\bibfnamefont {A.}~\bibnamefont {Liebsch}},
  \ and\ \bibinfo {author} {\bibfnamefont {S.}~\bibnamefont {Lounis}},\ }\href
  {\doibase 10.1103/PhysRevB.93.115123} {\bibfield  {journal} {\bibinfo
  {journal} {Phys. Rev. B}\ }\textbf {\bibinfo {volume} {93}},\ \bibinfo
  {pages} {115123} (\bibinfo {year} {2016})}\BibitemShut {NoStop}%
\bibitem [{\citenamefont {Choi}\ \emph {et~al.}(2017)\citenamefont {Choi},
  \citenamefont {Abufager}, \citenamefont {Limot},\ and\ \citenamefont
  {Lorente}}]{Choi:JCP:2017}%
  \BibitemOpen
  \bibfield  {author} {\bibinfo {author} {\bibfnamefont {D.-J.}\ \bibnamefont
  {Choi}}, \bibinfo {author} {\bibfnamefont {P.}~\bibnamefont {Abufager}},
  \bibinfo {author} {\bibfnamefont {L.}~\bibnamefont {Limot}}, \ and\ \bibinfo
  {author} {\bibfnamefont {N.}~\bibnamefont {Lorente}},\ }\href {\doibase
  10.1063/1.4972874} {\bibfield  {journal} {\bibinfo  {journal} {The Journal of
  Chemical Physics}\ }\textbf {\bibinfo {volume} {146}},\ \bibinfo {pages}
  {092309} (\bibinfo {year} {2017})}\BibitemShut {NoStop}%
\bibitem [{\citenamefont {Droghetti}\ and\ \citenamefont
  {Rungger}(2017)}]{Droghetti:PRB:2017}%
  \BibitemOpen
  \bibfield  {author} {\bibinfo {author} {\bibfnamefont {A.}~\bibnamefont
  {Droghetti}}\ and\ \bibinfo {author} {\bibfnamefont {I.}~\bibnamefont
  {Rungger}},\ }\href {\doibase 10.1103/PhysRevB.95.085131} {\bibfield
  {journal} {\bibinfo  {journal} {Phys. Rev. B}\ }\textbf {\bibinfo {volume}
  {95}},\ \bibinfo {pages} {085131} (\bibinfo {year} {2017})}\BibitemShut
  {NoStop}%
\bibitem [{\citenamefont {Abragam}\ and\ \citenamefont
  {Bleaney}(2012)}]{Abragam:book:2012}%
  \BibitemOpen
  \bibfield  {author} {\bibinfo {author} {\bibfnamefont {A.}~\bibnamefont
  {Abragam}}\ and\ \bibinfo {author} {\bibfnamefont {B.}~\bibnamefont
  {Bleaney}},\ }\href@noop {} {\emph {\bibinfo {title} {Electron Paramagnetic
  Resonance of Transition Ions}}}\ (\bibinfo  {publisher} {Oxford University
  Press},\ \bibinfo {address} {Oxford},\ \bibinfo {year} {2012})\BibitemShut
  {NoStop}%
\bibitem [{\citenamefont {Pruschke}\ and\ \citenamefont
  {Grewe}(1989)}]{Pruschke:ZPB:1989}%
  \BibitemOpen
  \bibfield  {author} {\bibinfo {author} {\bibfnamefont {T.}~\bibnamefont
  {Pruschke}}\ and\ \bibinfo {author} {\bibfnamefont {N.}~\bibnamefont
  {Grewe}},\ }\href {\doibase 10.1007/BF01311391} {\bibfield  {journal}
  {\bibinfo  {journal} {Z. Phys. B}\ }\textbf {\bibinfo {volume} {74}},\
  \bibinfo {pages} {439} (\bibinfo {year} {1989})}\BibitemShut {NoStop}%
\bibitem [{\citenamefont {Haule}\ \emph {et~al.}(2001)\citenamefont {Haule},
  \citenamefont {Kirchner}, \citenamefont {Kroha},\ and\ \citenamefont
  {W\"olfle}}]{Haule:PRB:2001}%
  \BibitemOpen
  \bibfield  {author} {\bibinfo {author} {\bibfnamefont {K.}~\bibnamefont
  {Haule}}, \bibinfo {author} {\bibfnamefont {S.}~\bibnamefont {Kirchner}},
  \bibinfo {author} {\bibfnamefont {J.}~\bibnamefont {Kroha}}, \ and\ \bibinfo
  {author} {\bibfnamefont {P.}~\bibnamefont {W\"olfle}},\ }\href {\doibase
  10.1103/PhysRevB.64.155111} {\bibfield  {journal} {\bibinfo  {journal} {Phys.
  Rev. B}\ }\textbf {\bibinfo {volume} {64}},\ \bibinfo {pages} {155111}
  (\bibinfo {year} {2001})}\BibitemShut {NoStop}%
\bibitem [{\citenamefont {Haule}\ \emph {et~al.}(2010)\citenamefont {Haule},
  \citenamefont {Yee},\ and\ \citenamefont {Kim}}]{Haule:PRB:2010}%
  \BibitemOpen
  \bibfield  {author} {\bibinfo {author} {\bibfnamefont {K.}~\bibnamefont
  {Haule}}, \bibinfo {author} {\bibfnamefont {C.-H.}\ \bibnamefont {Yee}}, \
  and\ \bibinfo {author} {\bibfnamefont {K.}~\bibnamefont {Kim}},\ }\href
  {\doibase 10.1103/PhysRevB.81.195107} {\bibfield  {journal} {\bibinfo
  {journal} {Phys. Rev. B}\ }\textbf {\bibinfo {volume} {81}},\ \bibinfo
  {pages} {195107} (\bibinfo {year} {2010})}\BibitemShut {NoStop}%
\bibitem [{\citenamefont {Jacob}\ and\ \citenamefont
  {Fern\'andez-Rossier}(2016)}]{Jacob:EPJB:2016}%
  \BibitemOpen
  \bibfield  {author} {\bibinfo {author} {\bibfnamefont {D.}~\bibnamefont
  {Jacob}}\ and\ \bibinfo {author} {\bibfnamefont {J.}~\bibnamefont
  {Fern\'andez-Rossier}},\ }\href {\doibase 10.1140/epjb/e2016-70402-2}
  {\bibfield  {journal} {\bibinfo  {journal} {The European Physical Journal B}\
  }\textbf {\bibinfo {volume} {89}},\ \bibinfo {pages} {210} (\bibinfo {year}
  {2016})}\BibitemShut {NoStop}%
\bibitem [{\citenamefont {Karan}\ \emph {et~al.}(2018)\citenamefont {Karan},
  \citenamefont {García}, \citenamefont {Karolak}, \citenamefont {Jacob},
  \citenamefont {Lorente},\ and\ \citenamefont {Berndt}}]{Karan:NL:2018}%
  \BibitemOpen
  \bibfield  {author} {\bibinfo {author} {\bibfnamefont {S.}~\bibnamefont
  {Karan}}, \bibinfo {author} {\bibfnamefont {C.}~\bibnamefont {García}},
  \bibinfo {author} {\bibfnamefont {M.}~\bibnamefont {Karolak}}, \bibinfo
  {author} {\bibfnamefont {D.}~\bibnamefont {Jacob}}, \bibinfo {author}
  {\bibfnamefont {N.}~\bibnamefont {Lorente}}, \ and\ \bibinfo {author}
  {\bibfnamefont {R.}~\bibnamefont {Berndt}},\ }\href {\doibase
  10.1021/acs.nanolett.7b03411} {\bibfield  {journal} {\bibinfo  {journal}
  {Nano Letters}\ }\textbf {\bibinfo {volume} {18}},\ \bibinfo {pages} {88}
  (\bibinfo {year} {2018})}\BibitemShut {NoStop}%
\bibitem [{\citenamefont {{Rubio-Verd{\'u}}}\ \emph {et~al.}(2017)\citenamefont
  {{Rubio-Verd{\'u}}}, \citenamefont {{Sarasola}}, \citenamefont {{Choi}},
  \citenamefont {{Majzik}}, \citenamefont {{Ebeling}}, \citenamefont {{Reyes
  Calvo}}, \citenamefont {{Ugeda}}, \citenamefont {{Garcia-Lekue}},
  \citenamefont {{S{\'a}nchez-Portal}},\ and\ \citenamefont
  {{Pascual}}}]{Rubio-Verdu::2017}%
  \BibitemOpen
  \bibfield  {author} {\bibinfo {author} {\bibfnamefont {C.}~\bibnamefont
  {{Rubio-Verd{\'u}}}}, \bibinfo {author} {\bibfnamefont {A.}~\bibnamefont
  {{Sarasola}}}, \bibinfo {author} {\bibfnamefont {D.-J.}\ \bibnamefont
  {{Choi}}}, \bibinfo {author} {\bibfnamefont {Z.}~\bibnamefont {{Majzik}}},
  \bibinfo {author} {\bibfnamefont {R.}~\bibnamefont {{Ebeling}}}, \bibinfo
  {author} {\bibfnamefont {M.}~\bibnamefont {{Reyes Calvo}}}, \bibinfo {author}
  {\bibfnamefont {M.~M.}\ \bibnamefont {{Ugeda}}}, \bibinfo {author}
  {\bibfnamefont {A.}~\bibnamefont {{Garcia-Lekue}}}, \bibinfo {author}
  {\bibfnamefont {D.}~\bibnamefont {{S{\'a}nchez-Portal}}}, \ and\ \bibinfo
  {author} {\bibfnamefont {J.~I.}\ \bibnamefont {{Pascual}}},\ }\href@noop {}
  {\enquote {\bibinfo {title} {{Orbital-selective spin excitation of a magnetic
  porphyrin}},}\ } (\bibinfo {year} {2017}),\ \Eprint
  {http://arxiv.org/abs/1708.01268} {1708.01268} \BibitemShut {NoStop}%
\bibitem [{\citenamefont {Ormaza}\ \emph {et~al.}(2017)\citenamefont {Ormaza},
  \citenamefont {Abufager}, \citenamefont {Verlhac}, \citenamefont
  {Bachellier}, \citenamefont {Bocquet}, \citenamefont {Lorente},\ and\
  \citenamefont {Limot}}]{Ormaza:NatComm:2017}%
  \BibitemOpen
  \bibfield  {author} {\bibinfo {author} {\bibfnamefont {M.}~\bibnamefont
  {Ormaza}}, \bibinfo {author} {\bibfnamefont {P.}~\bibnamefont {Abufager}},
  \bibinfo {author} {\bibfnamefont {B.}~\bibnamefont {Verlhac}}, \bibinfo
  {author} {\bibfnamefont {N.}~\bibnamefont {Bachellier}}, \bibinfo {author}
  {\bibfnamefont {M.-L.}\ \bibnamefont {Bocquet}}, \bibinfo {author}
  {\bibfnamefont {N.}~\bibnamefont {Lorente}}, \ and\ \bibinfo {author}
  {\bibfnamefont {L.}~\bibnamefont {Limot}},\ }\href {\doibase
  10.1038/s41467-017-02151-6} {\bibfield  {journal} {\bibinfo  {journal}
  {Nature Comm.}\ }\textbf {\bibinfo {volume} {8}},\ \bibinfo {pages} {1974}
  (\bibinfo {year} {2017})}\BibitemShut {NoStop}%
\bibitem [{\citenamefont {Schiller}\ and\ \citenamefont
  {De~Leo}(2008)}]{Schiller:PRB:2008}%
  \BibitemOpen
  \bibfield  {author} {\bibinfo {author} {\bibfnamefont {A.}~\bibnamefont
  {Schiller}}\ and\ \bibinfo {author} {\bibfnamefont {L.}~\bibnamefont
  {De~Leo}},\ }\href {\doibase 10.1103/PhysRevB.77.075114} {\bibfield
  {journal} {\bibinfo  {journal} {Phys. Rev. B}\ }\textbf {\bibinfo {volume}
  {77}},\ \bibinfo {pages} {075114} (\bibinfo {year} {2008})}\BibitemShut
  {NoStop}%
\bibitem [{\citenamefont {Schrieffer}\ and\ \citenamefont
  {Wolff}(1966)}]{Schrieffer:PR:1966}%
  \BibitemOpen
  \bibfield  {author} {\bibinfo {author} {\bibfnamefont {J.~R.}\ \bibnamefont
  {Schrieffer}}\ and\ \bibinfo {author} {\bibfnamefont {P.~A.}\ \bibnamefont
  {Wolff}},\ }\href {\doibase 10.1103/PhysRev.149.491} {\bibfield  {journal}
  {\bibinfo  {journal} {Phys. Rev.}\ }\textbf {\bibinfo {volume} {149}},\
  \bibinfo {pages} {491} (\bibinfo {year} {1966})}\BibitemShut {NoStop}%
\bibitem [{Note1()}]{Note1}%
  \BibitemOpen
  \bibinfo {note} {The excitations energies $\protect \mathaccentV
  {tilde}07E\Delta _0$ and $\protect \mathaccentV {tilde}07E\Delta _1$ are most
  conveniently extracted from the pseudo particle spectra associated to the
  corresponding excited many body states which show sharp peak features at the
  positions of the spin flip steps in the real electron spectrum, as shown in
  previous work.\cite {Jacob:EPJB:2016}}\BibitemShut {NoStop}%
\bibitem [{Note2()}]{Note2}%
  \BibitemOpen
  \bibinfo {note} {The $N+1=3$ subspace has spin-1/2 (one hole) and thus does
  not give rise to MA.}\BibitemShut {Stop}%
\end{thebibliography}%

\end{document}